\documentclass[showpacs,preprint,superscriptaddress,amsmath,amssymb]{revtex4-1}
\usepackage{txfonts}
\usepackage{graphicx}% Include figure files
\usepackage{dcolumn}% Align table columns on decimal point
\usepackage{bm}% bold math
\usepackage{epsfig}
\usepackage{booktabs}
\usepackage{subfigure}
\usepackage{graphics}
\usepackage{amssymb}
\usepackage{amsmath}
\usepackage{array}
\usepackage{color}
\usepackage{tabularx}
\usepackage{multirow}

\usepackage{xcolor}
\usepackage{algorithm}
\usepackage{algpseudocode}

\begin{document}
	
	%\preprint{APS/123-QED}
	
\title{A consistent and conservative diffuse-domain lattice Boltzmann method for multiphase flows in complex geometries}
	
\author{Xi Liu}
%\email[]{zhancj\_@hust.edu.cn}
\affiliation{School of Mathematics and Statistics, Huazhong University of Science and Technology, Wuhan, 430074, China}
\author{Chengjie Zhan}
%\email[]{zhancj\_@hust.edu.cn}
\affiliation{School of Mathematics and Statistics, Huazhong University of Science and Technology, Wuhan, 430074, China}
\author{Ying Chen}
%\email[]{zhancj\_@hust.edu.cn}
\affiliation{School of Mathematics and Statistics, Huazhong University of Science and Technology, Wuhan, 430074, China}
\author{Zhenhua Chai}
\email[Corresponding author: ]{hustczh@hust.edu.cn}
\affiliation{School of Mathematics and Statistics, Huazhong University of Science and Technology, Wuhan, 430074, China}
\affiliation{Institute of Interdisciplinary Research for Mathematics and Applied Science, Huazhong University of Science and Technology, Wuhan, 430074, China}
\affiliation{Hubei Key Laboratory of Engineering Modeling and Scientific Computing, Huazhong University of Science and Technology, Wuhan, 430074, China}
\author{Baochang Shi}
%\email[]{shibc@hust.edu.cn}
\affiliation{School of Mathematics and Statistics, Huazhong University of Science and Technology, Wuhan, 430074, China}
\affiliation{Institute of Interdisciplinary Research for Mathematics and Applied Science, Huazhong University of Science and Technology, Wuhan, 430074, China}
\affiliation{Hubei Key Laboratory of Engineering Modeling and Scientific Computing, Huazhong University of Science and Technology, Wuhan, 430074, China}
%\thanks{}

	\date{\today}% It is always \today, today,
	
	%%%%% Begin Abstract %%%%%%%%%%%
	\begin{abstract}
	Modeling and simulation of multiphase flows in complex geomerties are challenging due to the complexity in describing the interface topology changes among different phases and the difficulty in implementing the boundary conditions on the irregular solid surface. In this work, we first developed a diffuse-domain (DD) based phase-field model for multiphase flows in complex geometries. In this model, the irregular fluid region is embedded into a larger and regular domain by introducing a smooth characteristic function. Then, the reduction-consistent and conservative phase-field equation for the multiphase field and the consistent and conservative Navier-Stokes equations for the flow field are reformulated as the diffuse-domain based consistent and conservative (DD-CC) equations where some additional source terms are added to reflect the effects of boundary conditions. In this case, there is no need to directly treat the complex boundary conditions on the irregular solid surface, and additionally, based on a matched asymptotic analysis, it is also shown that the DD-CC equations can converge to the original governing equations as the interface width parameter tends to zero. Furthermore, to solve the DD-CC equations, we proposed a novel and simple lattice Boltzmann (LB) method with a Hermite-moment-based collision matrix which can not only keep consistent and conservation properties, but also improve the numerical stability with a flexible parameter. With the help of the direct Taylor expansion, the macroscopic DD-CC equations can be recovered correctly from the present LB method. Finally, to test the capacity of LB method, several benchmarks and complex problems are considered, and the numerical results show that the present LB method is accurate and efficient for the multiphase flows in complex geomerties.   
	\end{abstract}
	%%%%% end %%%%%%%%%%%
	
	%\pacs{44.05.+e, 02.70.-c}% PACS, the Physics and Astronomy
	
	\maketitle
	
\section{Introduction}
\label{sec:level1}
Multiphase flows in complex geometries are ubiquitous in nature and numerous industrial applications, such as liquid drops splashing on the ground, enhanced oil recovery and double emulsion production in microfluidics \cite{CHEN2015,Gan2009,Li2005}. In these problems, the dynamics of the fluid-fluid interface and the moving contact lines (MCLs) on irregular solid surface have a significant influence on the behavior of multiphase flows confined in complex geometries. Thus, in the modeling and simulation of such complex problems, how to accurately capture the fluid-fluid interface and treat the boundary conditions on irregular solid surface, especially the wetting boundary condition, are two core issues needed to be considered.

To describe the multiple interfacial dynamics, the phase-field method \cite{Lowengrub1998, Shen2011Modeling} where a diffusive interface with a small but finite width is adopted to replace the sharp interface between different phases, is commonly applied due to its capacity in capturing the topological changes implicitly. Based on the phase-field theory, the Cahn-Hilliard (CH) equation \cite{Boyer2014,Cahn1958,Dong2018} has been widely used in the study of the multiphase flows owing to its advantage in keeping the reduction-consistent property and thermodynamic consistency. However, the works on $N$-phase ($N>2$) flows in complex geometries are relatively rare, especially on how to deal with the MCLs on irregular solid surface efficiently.

\begin{figure}[H]
	\centering
	\includegraphics[width=2in]{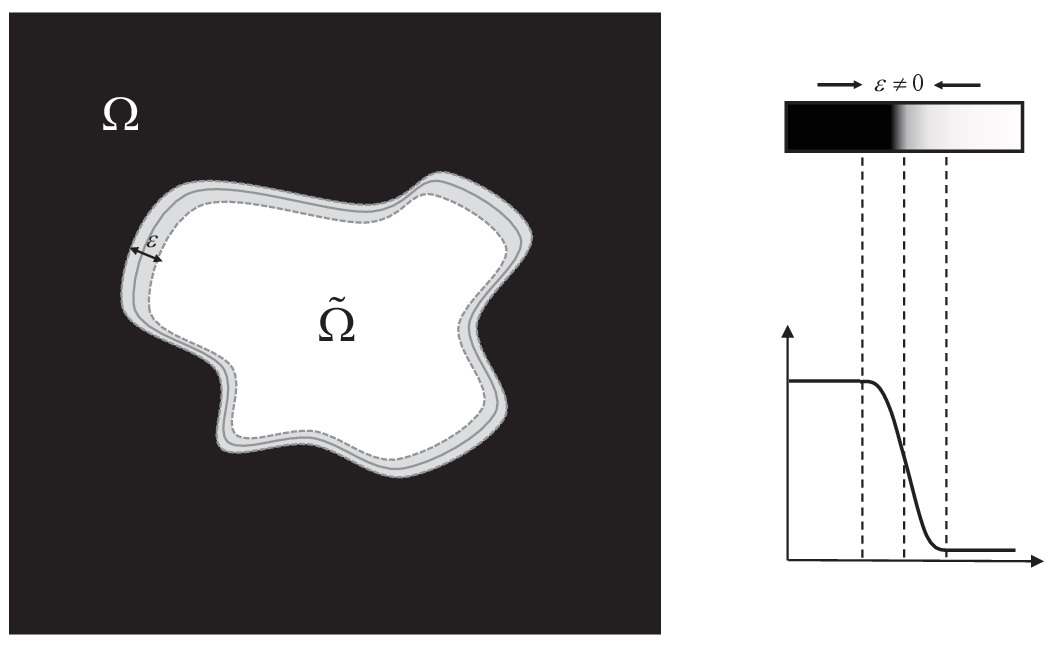}
	\caption{Schematic of the DD method.}
	\label{Exp1:Fig0}
\end{figure}

The diffuse-domain (DD) method  \cite{Aland2010TwophaseFI,Guo2020ADD,Li2009SOLVINGPI,liu2022,Yang2023b,Yang2023a,yu2020Higher}, as a diffuse interface approach, has received increasing attention in study of two-phase flows in complex geometries. As shown in Fig. \ref{Exp1:Fig0}, the basic idea of the DD method is that the complex fluid region is embedded into a larger and regular domain with a smooth characteristic function being used to enforce the boundary conditions at solid-fluid interface. Based on this idea, two main approaches have been developed to treat the boundary conditions on complex solid surface. The first one is based on a modified multi-component CH system, where one component is initially fixed as a solid phase, and the dynamics of fluid phases can be captured by solving the governing equations for remaining components \cite{Yang2023b,Yang2023a}. This approach has been used to capture two/ternary phase flows in arbitrary domains \cite{Yang2023b,Yang2023a}, but the CH system dose not satisfy the reduction-consistent property for multiphase flows \cite{Yang2023b}. Besides, this approach also suffers from another problem, i.e., an obvious shrinkage of the droplet, and it needs to be improved with an interfacial correction technique \cite{xia2022}. The second one is to directly reformulate the original governing equations at a larger and regular domain, which are also called the DD equations \cite{Li2009SOLVINGPI,yu2020Higher}. This approach establishes the DD equations in both the original and extended domains by using a phase-field parameter $\phi$ that varies continuously across the interface, and some additional source terms are added to enforce different boundary conditions. In addition, by using the matched asymptotic expansion, one can show that the DD equations would converge to the original governing equations and the corresponding boundary conditions on solid surface as the thickness of the DD interface $\varepsilon_0$ shrinks to zero. Actually, the second DD approach has been successfully utilized to investigate two-phase flows in complex geometries. For instance, Aland et al. \cite{Aland2010TwophaseFI} first coupled the DD method with the standard diffuse-interface method to study the two-phase flows in complex geometries. Later, Guo et al. \cite{Guo2020ADD} developed a thermodynamically consistent diffuse-interface model coupled with the DD method for the two-phase flows with large density ratio. Recently, Liu et al. \cite{liu2022} proposed a DD based consistent and conservative (DD-CC) phase-field model for two-phase flows in complex geometries, and investigated the effects of the wettability and viscosity ratio on the interfacial dynamics. To the best of our knowledge, however, there is no available DD model that can be used to treat $N$-phase ($N>2$) fluids in contact with complex solid surface, which is mainly caused by the difficulty in enforcing different boundary conditions of multiphase fluids.

To study $N$-phase flows in complex geometries, in this work, we develop a DD based consistent and conservative phase-field model for the first time, which can inherit the advantages of the phase-field method in capturing multiple interfacial dynamics implicitly and keeping the reduction-consistent property and thermodynamic consistency, and the DD method in treating boundary conditions on the irregular solid surface. Furthermore, different from the traditional numerical methods for the DD equations \cite{Aland2010TwophaseFI,Guo2020ADD}, here we focus on the mesoscopic lattice Boltzmann (LB) method \cite{Chai2020Multiple,Chai2023Multiple,Chen1998LATTICEBM,Krueger2016TheLB,Succi2001TheLB}
for its potential advantage in parallel scalability when incorporating complex physics models, and develop a new LB method to solve the DD-CC equations. It is worth noting that compared to classic multiple-relaxation-time (MRT)-LB method \cite{Dd2002}, the current LB (hereafter DD-CCLB) method contains the Hermite-monent based collision matrix \cite{Chai2023Multiple,Coreixas2019, Krueger2016TheLB} with a flexible parameter $d_0$ \cite{Chai2023Multiple}, which can be used to improve the numerical stability. The rest of this paper is organized as follows. In section 2, the DD-CC equations for the $N$-phase flows in complex geometries are first proposed, followed by the devloped DD-CCLB method in section 3. In section 4, we conduct some simulations to test the present DD-CCLB method, and finally, some conclusions are given in section 5.

\section{Mathematical equations}
\label{sec:level2}
\subsection{\label{sec:level2.1} The reduction-consistent Cahn-Hilliard equation for multiphase flows}
Based on the phase-field theory for $N$-phase ($N\geq 2$) incompressible and immiscible fluids in an irregular domain $\tilde{\Omega}$, one can define the total mixture free energy $\mathcal{F}(\vec{C}, \nabla \vec{C})$ for $N$-phase system \cite{Boyer2014,Dong2018} as
\begin{equation}
	\mathcal{F}(\vec{C}, \nabla \vec{C})=\int\left[\mathcal{F}_0(\vec{C})+\sum_{i, j=1}^N \frac{\kappa_{i j}}{2} \nabla C_i \cdot \nabla C_j\right] d \tilde{\Omega},
	\label{eq:2.1}
\end{equation}
where $C_i$ represents the volume fraction of $i$th fluid with the following constraint:
\begin{equation}
	\sum_{i=1}^N C_i=1, \quad 0\leq C_i\leq 1.
\end{equation}
In Eq. (\ref{eq:2.1}), the first term on the right-hand side $\mathcal{F}_0(\vec{C})$ is the bulk free energy, and it can be given by $\mathcal{F}_0(\vec{C})=\sum_{i, j=1}^N \beta_{i j}\left[g\left(C_i\right)+g\left(C_j\right)-g\left(C_i+C_j\right)\right]$  with $\vec{C}=\left(C_1, C_2, \ldots, C_N\right)$ and $g(C)=C^2(C-1)^2$. The second term is the excess free energy at the interfacial region. Here $\beta_{i j}$ and $\kappa_{i j}$ are two constant parameters related to the surface tension $\sigma_{ij}$ and the interface width $\varepsilon$, and can be expressed as $\beta_{i j}=3{\varepsilon}/\sigma_{ij}$ and $\kappa_{i j}=-3\varepsilon\sigma_{ij}/4$. In addition, the surface tension $\sigma_{ij}$ has the symmetric property, i.e., $\sigma_{ij}=\sigma_{ji}$, $\sigma_{ji}>0$ with $i\neq j$, and $\sigma_{ii}=0$.
By minimizing the total mixture free energy, we can derive the following expression of the chemical potential,
\begin{equation}
	\mu_{C_i}=\frac{\delta \mathcal{F}} { \delta C_i}=	\partial_{C_i}\mathcal{F}_0	-\sum_{ j=1}^N\kappa_{i j}\nabla ^2 C_j.
\end{equation}
Then the CH equation for multiphase flows can be written as \cite{Dong2018}
\begin{equation}
	\partial_{t}C_i+\nabla\cdot(C_i\textbf{u})=\sum_{j=1}^N\nabla\cdot(M_{ij}\nabla\mu_{C_j}),
	\label{eq:2.4}
\end{equation}
where $M_{i j}$ is the mobility. To guarantee the reduction-consistent property, which means when the $i$th fluid (or more fluids) is absent in $N$-phase system, the CH equation for $N$-phase flows should reduce to the one containing $M$-phase flows ($M<N$), the mobility $M_{i j}$ and $C_i$ are required to be zero when the $i$th fluid is absent. To this end, here we adopt the following expression of mobility $M_{i j}$,
\begin{equation}
	M_{i j}=
	\begin{cases}
		-m_0C_iC_j, & i\neq j, \\
		-\sum_{j, j \neq i} M_{i j},& i= j. 
	\end{cases}
	\label{eq:2.5}
\end{equation}
In addition, this type of $M_{i j}$ could remedy a major defect of droplet shrinkage because the maximum principle property can be preserved in this reduction-consistent equation \cite{Acosta-Soba2023,Elliott1996}. 

When the multiphase fluids are in contact with the solid surface, the wetting boundary condition should be used to account for the effect of the contact angles formed by fluid-fluid interface and solid wall on the interface dynamics. In this work, the following wetting boundary condition with a reduction-consistent property for $N$-phase immiscible fluids is adopted \cite{Dong2017},
\begin{equation}
	\boldsymbol{n}_w \cdot \nabla C_i=\sum_{j=1}^N \xi_{i j} C_i C_j,  \quad 1 \leq i \leq N.
	\label{eq:2.7}
\end{equation}
Here $\boldsymbol{n}_w$ is the normal vector of solid wall, and the explicit expression of $\xi_{i j}$ can be given by
\begin{equation}
	\xi_{i j}=
	\begin{cases}
		\frac{4}{\varepsilon}\cos\theta_{ij}, &1 \leq i \neq j \leq N, \\
		0,& 1\leq i=j \leq N,
	\end{cases}
\end{equation}
where $\theta_{ij}$  is the contact angle formed by fluid-fluid interface of $i, j$ phases and solid wall, and is measured on the side of the $i$th fluid. Additionally, we also introduce $\theta_{iN}$ ($1 \leq i  \leq N-1$) to denote the contact angle between solid wall and fluid-fluid interface formed by the $i$th and $N$th phases, and $N-1$ independent contact angles $\theta_{iN}$ can be used to calculate $\theta_{ij}$ by \cite{Dong2017},
\begin{equation}
	\cos \theta_{i j}=\left(\frac{\sigma_{i N}}{\sigma_{i j}} \cos \theta_{i N}-\frac{\sigma_{j N}}{\sigma_{i j}} \cos \theta_{j N}\right).
	\label{eq:2.9}
\end{equation}		

\subsection{\label{sec:level2.2} The conservative and consistent Navier-Stokes equations for flow field}
For $N$-phase immiscible and incompressible flows in the irregular domain ${\tilde{\Omega}}$, the Navier-Stokes (NS) equations can be written as \cite{Dong2018,Huang2020AConsistent}
\begin{subequations}
	\begin{equation}
		\nabla \cdot\textbf{u}=0,
		\label{eq:2.2a}
	\end{equation}
	\begin{equation}
		\partial_{t}(\rho\textbf{u})+\nabla \cdot(\textbf{m}\textbf{u})=-\nabla P+\nabla\cdot[\rho\nu (\nabla\textbf{u}+\nabla\textbf{u}^{T})]+\textbf{F}_{s},
		\label{eq:2.2b}
	\end{equation}
	\label{eq:2.2}
\end{subequations}
where $\rho$ is the fluid density, $\textbf{m}$ is the mass flux, $\nu$ is the kinematic viscosity, $P$ is the pressure, $\textbf{F}_{s}=\sum_{i}\mu_{C_i}\nabla C_i$ is the surface tension force.
In the multiphase system, the density and viscosity are usually assumed to be a linear function of the phase-field variable $C_i$,
\begin{equation}
	\rho=\sum_{i}C_i\rho_i,\quad\nu=\sum_{i}C_i\nu_i,
\end{equation}
where $\rho_i$ and $\nu_i $ are the density and viscosity of the $i$th-phase fluid, respectively. To guarantee the consistency of reduction, the consistency of mass and momentum transport, and the consistency of mass conservation \cite{Huang2020AConsistent}, the mass flux $\textbf{m}$ should be designed as 
\begin{equation}
	\textbf{m}=\rho\mathbf{u}+ \textbf{m}^C,
\end{equation}
with
\begin{equation}
	\textbf{m}^C=-\sum_{ij}\rho_iM_{ij}\nabla\mu_{C_j}.
\end{equation}
Here $ \textbf{m}^C$ denotes the mass diffusion between different phases. In addition, the no slip boundary condition is imposed on the solid surface. 

\subsection{\label{sec:level2.3}The DD equations for multiphase flows in complex geometries}
In this section, the DD method is applied to reformulate CC equations (\ref{eq:2.4}, \ref{eq:2.2}) coupled with the boundary conditions in a larger and regular domain $\Omega$ 
\cite{Aland2010TwophaseFI,Guo2020ADD, Li2009SOLVINGPI,liu2022,yu2020Higher}. Through introducing a smooth function $\phi$ to label the diffuse-interface between the fluid and the solid surface, the following DD-CC equations can be obtained 
\begin{subequations}
	\begin{equation}
		\partial_{t}(\phi C_i)+\nabla\cdot(\phi C_i\textbf{u})=\sum_{j=1}^N\nabla\cdot(\phi M_{ij}\nabla\mu_{C_j}),
		\label{eq:2.13a}
	\end{equation}
	\begin{equation}
		\nabla \cdot(\phi\textbf{u})=0,
		\label{eq:2.13b}
	\end{equation}
	\begin{equation}
		\begin{aligned}
			\partial_{t}(\phi\rho\textbf{u})+\nabla \cdot(\phi\rho\textbf{u}\textbf{u}+\phi \textbf{m}^{C}\textbf{u})=-\phi\nabla P+\nabla\cdot[\phi\rho \nu (\nabla\textbf{u}+\nabla\textbf{u}^{T})]+\sum_{i=1}^{N}\phi\mu_{C_i}\nabla C_i\\
			-\frac{(1-\phi)}{\varepsilon_0^3}\textbf{u},		
		\end{aligned}
		\label{eq:2.13c}
	\end{equation}
	\label{eq:2.13}
\end{subequations}
with
\begin{equation}
	\begin{aligned}
		\phi\mu_{C_i}=\sum_{j=1}^N 2 \beta_{ij}\phi \left[g^{\prime}\left(C_i\right)-g^{\prime}\left(C_i+C_j\right)\right]+\sum_{j=1}^N	\frac{3\varepsilon}{4}\sigma_{ij} \nabla\cdot(\phi \nabla C_j)\\
		+\sum_{j=1}^N 3\sigma_{ij}\sum_{q\neq j}\cos\theta_{jq}C_j C_q|\nabla\phi|,
	\end{aligned}
	\label{eq:2.14}
\end{equation}
where $\phi=1$ in the original complex domain $\tilde{\Omega}$, while $\phi=0$ in $\Omega/\tilde{\Omega}$, and $\phi=1/2$ is used to mark the solid boundary $\partial \tilde{\Omega}$. $\varepsilon_0$ is the thickness of the diffuse layer. In the larger and regular domain $\Omega$, the boundary conditions, $\textbf{u}=0$ , $\textbf{n}\cdot\nabla C_i=0$, and $\textbf{n}\cdot\nabla \mu_{C_i}=0$, are imposed on $\partial{\Omega}$.
It is worth mentioning that there is no need to directly handle the complex boundary conditions on $\partial \tilde\Omega$. Additionally, with the help of the matched asymptotic analysis, the DD-CC equations can asymptotically converge to the original equations as $\varepsilon_0 \rightarrow 0$. Here we only take the DD-CH equation (\ref{eq:2.13a}) as an example, and show the details in Appendix \ref{sec:appendixA}.
\section{\label{sec:level3}The LB method for DD equations}
In this section, we develop a novel and simple LB method for the DD-CC equations (\ref{eq:2.13}). Due to the fact that the mobility $M_{ij}$ (\ref{eq:2.5}) is related to $C_i$, the standard LB method would be unstable when it adopted to the DD-CH equation. To overcome this problem, the diffusion flux $\phi M_{ii}\nabla\mu_{C_i}$ is divided into two parts, $\phi m_0\nabla\mu_{C_i}$ and $\left(\phi M_{ii}-\phi m_0\right)\nabla\mu_{C_i}$, in which the first part containing a constant mobility $m_0$ can be seen as a diffusion term, and the other part can be handled as a source term. Then we propose a new Hermite-moment-based LB method for the DD-CC equations (\ref{eq:2.13}),
\begin{equation}
	\begin{aligned}
		f^{i}_{p}\left(\mathbf{x}+\mathbf{c}_p \delta t, t+\delta t\right)=f^{i}_{p}(\mathbf{x}, t)-\Lambda^{f^{i}}_{pq}\left[f^{i}_{q}(\mathbf{x}, t)-f_{q}^{i,eq}(\mathbf{x}, t)\right]\\
		+\delta t\left(\delta_{pq}-\frac{\Lambda^{f^{i}}_{pq}}{2}\right) F^{i}_{q}(\mathbf{x}, t),
	\end{aligned}
	\label{eq:3.1}
\end{equation}
\begin{equation}
	\begin{aligned}
		g_{p}\left(\mathbf{x}+\mathbf{c}_p \delta t, t+\delta t\right)=g_{p}(\mathbf{x}, t)-\Lambda^g_{pq}\left[g_{q}(\mathbf{x}, t)-g_{q}^{e q}(\mathbf{x}, t)\right]\\
		+\delta t\left(\delta_{pq}-\frac{\Lambda^g_{pq}}{2}\right) G_{q}(\mathbf{x}, t),
	\end{aligned}
	\label{eq:3.2}
\end{equation}
where $f^{i}_{p}$ and $g_{p}(\textbf{x},{t})$ ($p=0,1...,k-1$) are the distribution functions for the $i$th-phase field and flow field with $k$ being the number of the discrete velocities, $f_{p}^{i,eq}(\textbf{x},{t})$ and $g_{p}^{eq}(\textbf{x},{t})$ are the corresponding equilibrium distribution functions, ${F}^{i}_{p}(\textbf{x},{t})$ and ${G}_{p}(\textbf{x},{t})$ are the distribution functions of the source and force terms. $\boldsymbol\Lambda^{f^i}$ and $\boldsymbol\Lambda^{g}$ are the $k\times k$ invertible collision matrices with the following forms,
\begin{equation}
	\boldsymbol\Lambda^{f^i}=\textbf{H}^{-1}\textbf{S}^{f^i}\textbf{H}, \quad \boldsymbol\Lambda^{g}=\textbf{H}^{-1}\textbf{S}^{g}\textbf{H}, 
\end{equation}
where $\textbf{H}$ is the Hermite-moment-based transformation matrix \cite{Chai2023Multiple,Coreixas2019,Krueger2016TheLB}, $\textbf{S}^{f^i}$ and $\textbf{S}^{g}$ are the relaxation matrices. To recover DD-CH equation (\ref{eq:2.13a}), the distribution functions $f_{p}^{i,eq}(\textbf{x},{t})$ and ${F}^{i}_{p}(\textbf{x},{t})$ should be designed properly, and can be given by
\begin{equation}
	f_{p}^{i,e q}(\mathbf{x}, t)= \begin{cases}\left(\omega_p-1\right) \eta_i \phi \mu_{C_i}+\phi C_i, & p=0, \\ \omega_p \eta_i \phi \mu_{C_i}+\omega_p \frac{\mathbf{c}_p \cdot \phi C_i \mathbf{u}}{c_s^2}, & p\neq 0,\end{cases}
	\label{eq:3.5}
\end{equation}
and 
\begin{equation}
	F^{i}_{p}(\mathbf{x}, t)=\omega_p \frac{\mathbf{c}_p\cdot\left[\partial_t(\phi C_i\mathbf{u})+c_s^2 \eta_i (\mu_{C_i} \nabla \phi-\textbf{S}_i)\right]}{c_s^2},
	\label{eq:3.6}
\end{equation}
where 
\begin{equation}
	\textbf{S}_i=\frac{\phi M_{ii}-\phi m_{0}}{m_{0}}\nabla\mu_{C_i}+\sum_{j\neq i}\frac{\phi M_{ij}}{m_0}\nabla\mu_{C_j}.
\end{equation}
Here $\mathbf{c}_p$ is the discrete velocity, $c_{s}$ is the lattice sound speed, $\omega_{p}$ is the weight coefficient satisfying $\sum\omega_{p}=1$. Usually the weight coefficient $\omega_{p}$ is fixed in MRT-LB method, but in the present LB method, $\omega_{p}$ is related to a flexible parameter $d_0$, which can be adjusted to improve the numerical stability and accuracy \cite{Chai2023Multiple}. To recover DD-NS equations (\ref{eq:2.13b} and \ref{eq:2.13c}), the distribution functions $g_{p}^{eq}(\textbf{x},{t})$ and ${G}_{p}(\textbf{x},{t})$ are designed as
\begin{equation}
	g_p^{e q}(\mathbf{x}, t)= \begin{cases}\left(\omega_p-1\right) \frac{\phi P}{c_s^2}+\rho_0+s_p, & p=0, \\
		\omega_p \frac{\phi P}{c_s^2}+s_p, & p \neq 0,\end{cases} 
	\label{eq:3.8}
\end{equation}
and
\begin{equation}
	G_p(\mathbf{x}, t)=\omega_p\left[\phi \mathbf{u} \cdot \nabla \rho +\frac{\mathbf{c}_p \cdot \mathbf{F}}{c_s^2}+\frac{M^{2G}_{\alpha{\alpha}}\left({c}_{p\alpha}{c}_{p\alpha}-c_s^2 \right)}{c^2 c_s^2-c_s^4}+\frac{M^{2G}_{\alpha\bar{\alpha}}{c}_{p\alpha}{c}_{p\bar{\alpha}}}{c_s^4}\right],
	\label{eq:3.10}
\end{equation}
where
\begin{equation}
	\begin{aligned}
		s_p=\omega_p\left[\frac{\mathbf{c}_p\cdot \phi \rho \mathbf{u}}{c_s^2}+\frac{\left(\phi \rho u_{\alpha}u_{\alpha}+\phi {m}^c_{\alpha} u_{\alpha}\right)\left({c}_{p\alpha}{c}_{p\alpha}-c_s^2 \right)}{c^2 c_s^2-c_s^4}\right.\\
		\left.+\left(\phi \rho u_{\alpha}u_{\bar{\alpha}}+\frac{\phi {m}^c_{\alpha} u_{\bar{\alpha}}+\phi u_{\alpha}{m}^c_{\bar{\alpha}} }{2}\right)\frac{{c}_{p\alpha}{c}_{p\bar{\alpha}}}{c_s^4}\right],
	\end{aligned}
\end{equation}
\begin{equation}
	\mathbf{M}^{2G}=\partial_t\left(\frac{\phi\mathbf{m}^{ C} \mathbf{u}+\phi\mathbf{u m}^{ C}}{2}\right)+c_s^2\left[\mathbf{u} \nabla(\phi \rho)+\nabla( \phi\rho) \mathbf{u}\right]+(c^2-3c_s^2)\mathbf{u} \cdot\nabla(\phi\rho) \mathbf{I},
\end{equation}
and
\begin{equation}
	\textbf{F}=-\nabla\cdot\left(\frac{\phi\textbf{m}^c\textbf{u}-\phi\textbf{u}\textbf{m}^c}{2}\right)+\sum_{i=1}^{N}\phi\mu_{C_i}\nabla C_i+P\nabla\phi-\frac{(1-\phi)}{\varepsilon_0^3}\textbf{u},
\end{equation}
where $\alpha$ represents the dimension of space and $\bar{\alpha}$ denotes all subscripts except $\alpha$. Through the direct Taylor analysis (see Appendix \ref{sec:appendixB} for details), the DD-CC equations can be recovered correctly with the following relations,
\begin{equation}
	m_0=\left(\frac{1}{s^{f^i}_{1}}-\frac{1}{2}\right)\eta_{i} c_{s}^{2}\delta {t},
	\label{eq:17}
\end{equation}
\begin{equation}
	\nu=\left(\frac{1}{s^{g}_{{21}}}-\frac{1}{2}\right)\frac{c^{2}-c_{s}^{2}}{2}\delta {t}, \quad
	\nu=\left(\frac{1}{s^{g}_{{22}}}-\frac{1}{2}\right)c_{s}^{2}\delta {t},
	\label{eq:16}
\end{equation}
where $s^{f^i}_{1}$, $s^{g}_{21}$ and $s^{g}_{{22}}$ are the relaxation parameters appeared in $\textbf{S}^{f^i}$ and $\textbf{S}^{g}$, respectively. $\eta_i$ is an adjustable parameter which can be applied to improve the numerical stability for a specified $m_0$ \cite{liu2022}. In addition, the macroscopic volume fraction parameter $C_i$, fluid velocity $\textbf{u}$, and pressure $P$ are calculated by
\begin{equation}
	\phi C_i=\sum_{p}f^i_{p},
\end{equation}
\begin{equation}
	\phi\rho\textbf{u}=\sum_{p}{\textbf{c}_{p}g_{p}}+\frac{1}{2}\delta {t}\textbf{F},
\end{equation}

\begin{equation}
	\begin{aligned}
		\phi P=\frac{c_s^2}{1-\omega_0}\left\{\sum_{p \neq 0} g_p+\left[\frac{1}{2}+\frac{K(1-d_0)(2-s^g_0)}{2s^g_0}\right] \delta t \phi\mathbf{u} \cdot \nabla \rho+ \delta t\frac{K}{2c^2} \partial_t\left(\phi\mathbf{m}^{\phi C} \cdot \mathbf{u}\right)\right.\\
		\left.+\delta t\frac{H}{s^g_{21}}  \mathbf{u} \cdot \nabla (\phi\rho)+s_0\right\},
	\end{aligned}
	\label{eq:19}
\end{equation}
where $K=1-d_0$ and $H=K(1-2d_0)(s^g_{21}-2)$ for two dimensional problems while $K=1-2d_0$ and $H=K(3-7d_0)(s^g_{21}-2)/2$ for three-dimensional problems. In our simulations, the D2Q9 lattice model is adopted for both phase and flow fields in two-dimensional problems, while in three-dimensional problems, the D3Q7 and D3Q15 lattice models are used for phase and flow fields, respectively. In the following, the discrete velocity $\mathbf{c}_p$, the weight coefficient $\omega_p$, and the relaxation matrix $\mathbf{S}$ in the Hermite-moment based lattice models are given by \cite{Chai2023Multiple}

D2Q9:
\begin{equation}
	\begin{aligned}
		\mathbf{c}_p&= \begin{cases}
			(0,0,0), & p=0, \\ 
			( \pm c,0),(0, \pm c), & p=1-4, \\
			( \pm c,\pm c), &p=5-8, \\		
		\end{cases}
		\quad 
		\omega_p=\begin{cases}
			1-2d_0+d_0^2, & p=0, \\
			(d_0-d_0^2)/2, &p=1-4,\\
			d_0^2/4, & p=5-8,\\
		\end{cases}\\
		\mathbf{S}&=\operatorname{diag}\left(s_0, s_1, s_1,s_{21},s_{21},s_{22},s_3,s_3,s_4\right).
	\end{aligned}
\end{equation}

D3Q7:
\begin{equation}
	\begin{aligned}
		\mathbf{c}_p&= \begin{cases}
			(0,0,0), & p=0 ,\\ 
			( \pm c,0,0),(0, \pm c,0),(0,0, \pm c), & p=1-6, \\
		\end{cases}
		\quad 
		\omega_p=\begin{cases}
			1-3d_0, & p=0, \\
			d_0/2, & p=1-6,\\
		\end{cases}\\
		\mathbf{S}&=\operatorname{diag}\left(s_0, s_1, s_1,s_{1},s_{21},s_{21},s_{21}\right).
	\end{aligned}
\end{equation}

D3Q15:
\begin{equation}
	\begin{aligned}
		\mathbf{c}_p&=\begin{cases}
			(0,0,0), & p=0 ,\\ 
			( \pm c,0,0),(0, \pm c,0),(0,0, \pm c), & p=1-6, \\
			( \pm c,\pm c,\pm c), & p=7-14, \\
		\end{cases}
		\omega_p=\begin{cases}
			1-3d_0+2d_0^2, & p=0, \\
			(d_0-d_0^2)/2, & p=1-6,\\
			d_0^2/8, & p=7-14,\\		
		\end{cases}\\
		\mathbf{S}&=\operatorname{diag}\left(s_0, s_1, s_1,s_{1},s_{21},s_{21},s_{21},s_{22},s_{22},s_{22},s_{3},s_{3},s_{3},s_{3},s_{4}\right),
	\end{aligned}
\end{equation}
where $d_0=c_s^2/c^2$ and $c=\delta x/\delta t$. 
\section{\label{sec:level4}Numerical results and discussion}
In this section, four benchmark problems, i.e., the spreading of a single droplet, the spreading of a compound droplet in ternary system, the spreading of a compound droplet in quaternary system on an ideal wall in two-dimensional space, and a compound droplet spreading on a solid sphere in three-dimensional space, are first used to validate the present DD-CCLB method. Then a complex problem, a compound droplet passing through an irregular channel, is applied to demonstrate the capacity of the DD-CCLB method for multiphase flows in complex geometries.
\subsection{\label{sec:level4.1}The spreading of a single droplet on an ideal wall}
\begin{figure}[H]
	\begin{center}
		\subfigure[]{ \label{fig1:subfig:60}
			\includegraphics[width=1.5in]{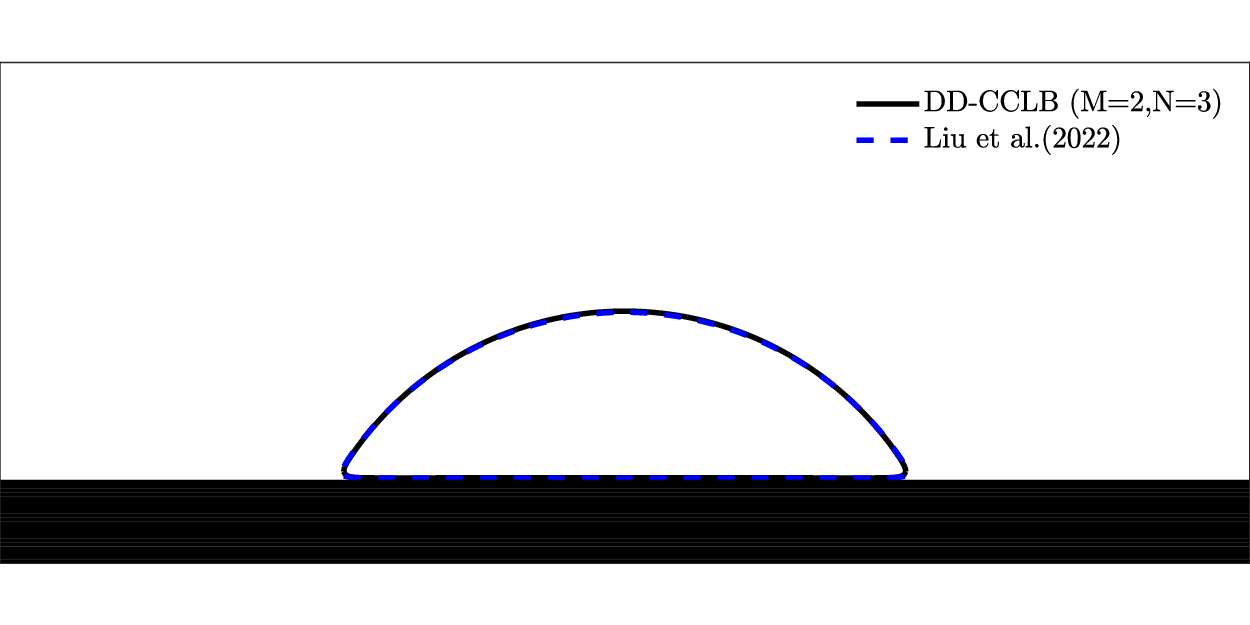}}
		\subfigure[]{ \label{fig1:subfig:90}
			\includegraphics[width=1.5in]{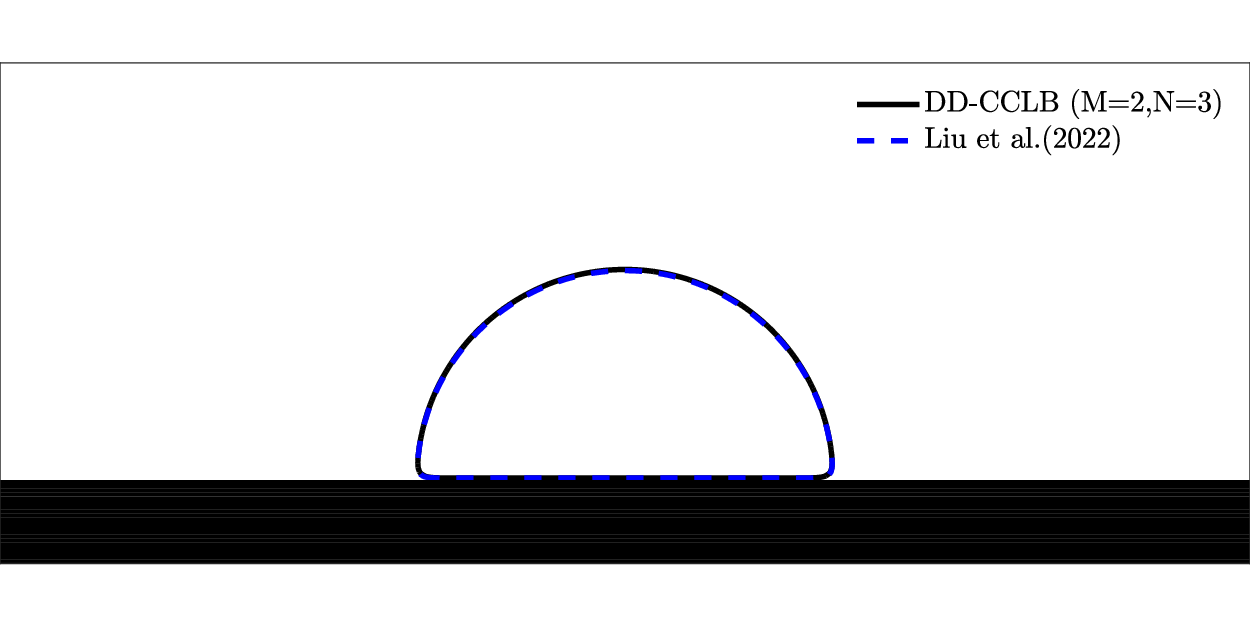}}
		\subfigure[]{ \label{fig1:subfig:120}
			\includegraphics[width=1.5in]{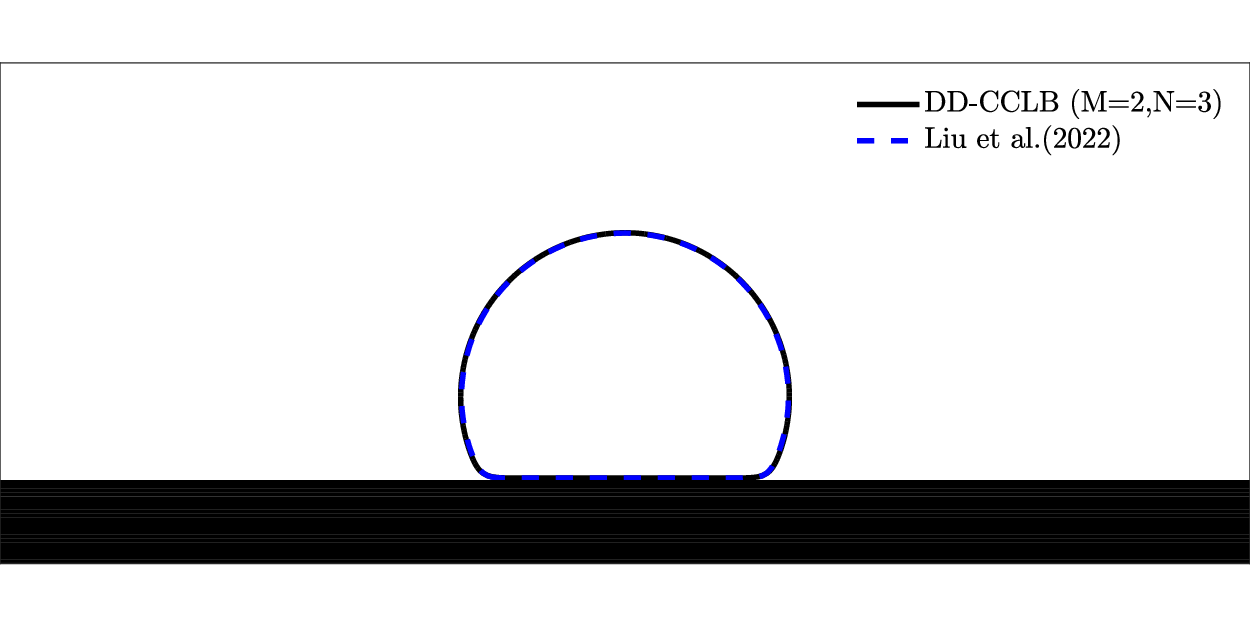}}
		\caption{The predicted equilibrium shape of a single droplet [The solid and dashed lines ($C=0.5$) are obtained by the present DD-CCLB method ($M=2$ and $N=3$) and two-phase method \cite{liu2022} ($M=N=2$) under conditions of prescribed contact angles: (a) $\theta=60^\circ$, (b) $\theta=90^\circ$, (c) $\theta=120^\circ$].}
	\end{center}
	\label{fig:ex1}
\end{figure}

We first consider the spreading of a single droplet on an ideal wall, which can be used as a good benchmark to test the reduction-consistent property of the DD-CCLB method. In this numerical experiment, we focus on a ternary-phase system with two fluids ($M = 2$ and $N = 3$), and the volume fraction of the third phase ($C_3$) is set to be zero. Initially, a semicircular droplet with the radius $R=50\delta x$ is deposited on the ideal wall, and some parameters are fixed as $\rho_1=1$, $\rho_2=\rho_3=0.01$, $\nu_1=\nu_2=\nu_3=0.1$, $\sigma_{ij}=0.001$ where $i\neq j$, $m_0=0.01$, $\varepsilon_0=0.16$, $\varepsilon=5\delta x$, $\delta x=0.04$, $c=20$, $d_0=0.4$,  $s^{fi}_{0}=1$, $s^{fi}_{1}=10/19$, $s^{fi}_{21}=s^{fi}_{22}=1.2$,  $s^{fi}_{3}=s^{fi}_{4}=1.2$,  $s^{g}_{0}=1$, $s^{g}_{1}=s^{g}_{3}=s^{g}_{4}=1.2$, and the boundary conditions are the same as those in Ref. \cite{liu2022}. As seen from Fig. \ref{fig:ex1}, the single droplet can form different steady patterns under different values of the contact angle $\theta$, and additionally, the present results are also in good agreement with those obtained by the two-phase method ($M=N=2$) \cite{liu2022}. Moreover, we give a quantitative comparison of the contact angle in Table \ref{table1}, and a good agreement between two different methods is observed, which indicates the present DD-CCLB method keeps the reduction-consistent property well.
\begin{table}
	\centering
	\caption{A comparison of the present DD-CCLB method ($M=2$ and $N=3$) and two-phase method \cite{liu2022} ($M=N=2$) in predicting the contact angle of a single droplet.}
	\begin{tabular}{ccccccc}
		\toprule
		Contact angles    && $60^{\circ}$ && $90^{\circ}$ && $120^{\circ}$  \\
		\midrule
		DD-CCLB      && $60.81^{\circ}$ && $90.29^{\circ}$ && $120.65^{\circ}$  \\
		Liu et al. \cite{liu2022}  && $60.85^{\circ}$ && $89.77^{\circ}$ && $120.72^{\circ}$  \\
		\bottomrule
	\end{tabular}
	\label{table1}
\end{table}

\subsection{\label{sec:level4.2}The spreading of a compound drop in ternary system on an ideal wall}
\begin{figure}[H]
	\begin{center}
		\subfigure[]{
			\includegraphics[width=1.50in]{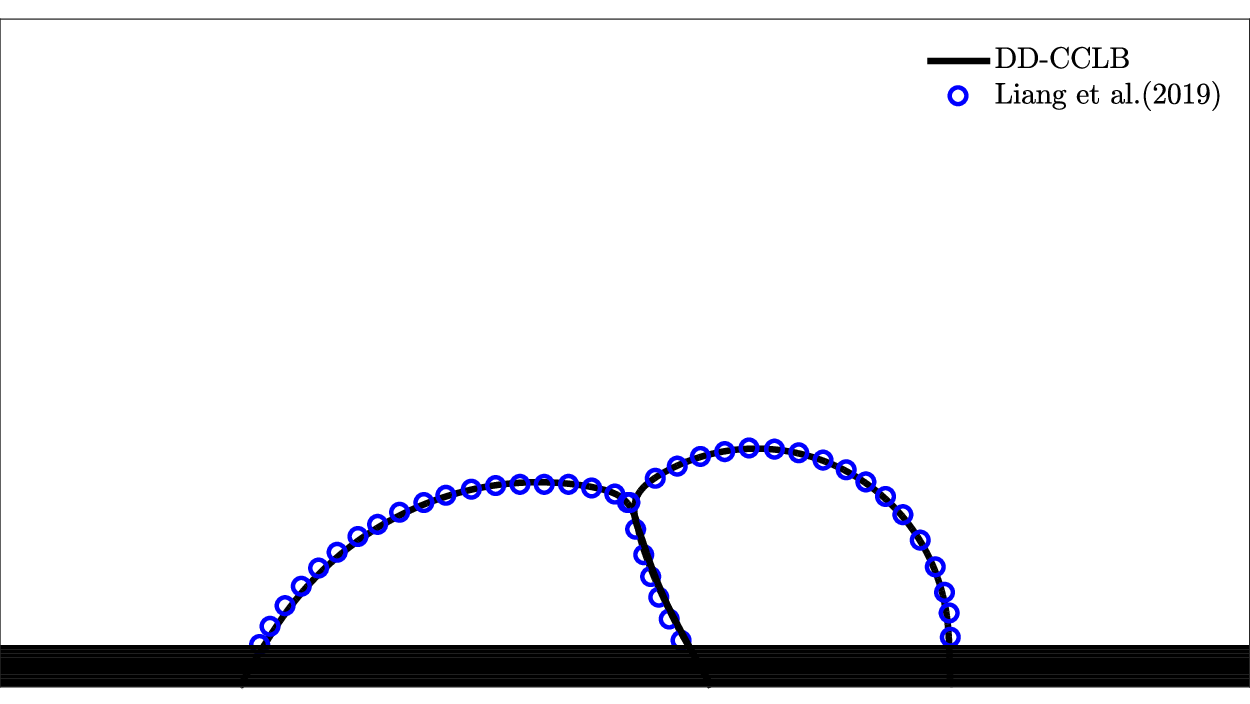}}
		\subfigure[]{ 
			\includegraphics[width=1.50in]{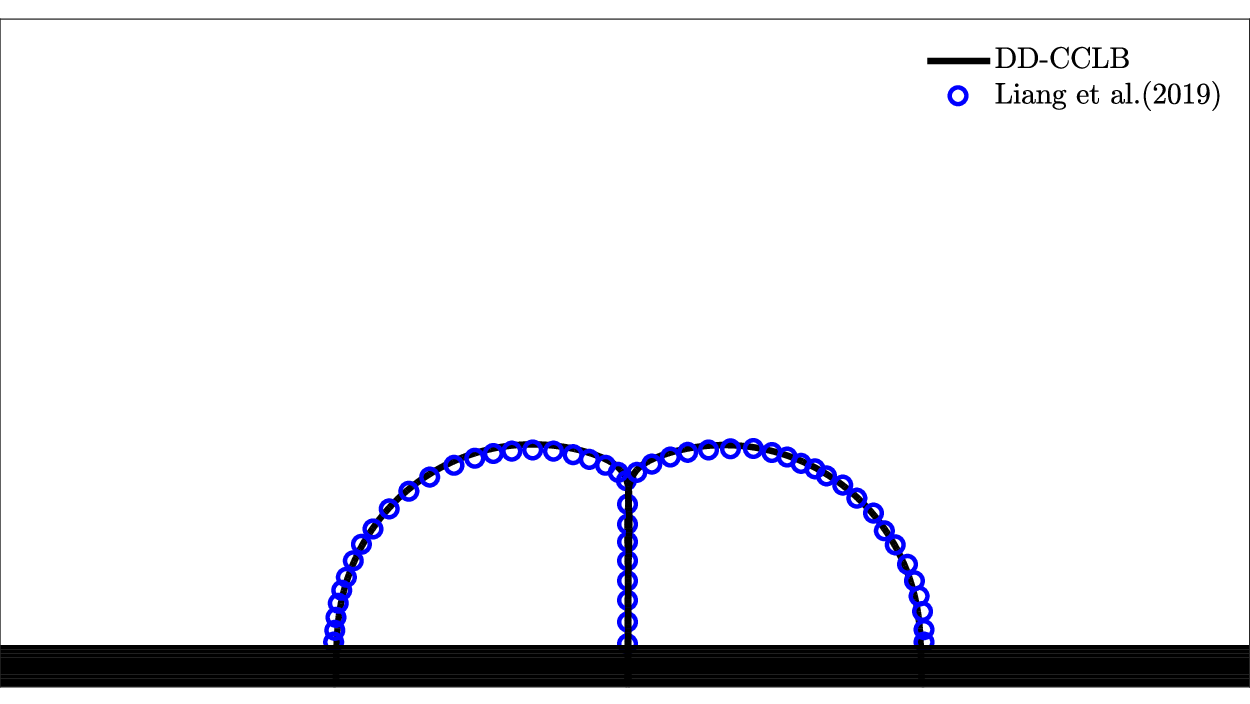}}
		\subfigure[]{ 
			\includegraphics[width=1.50in]{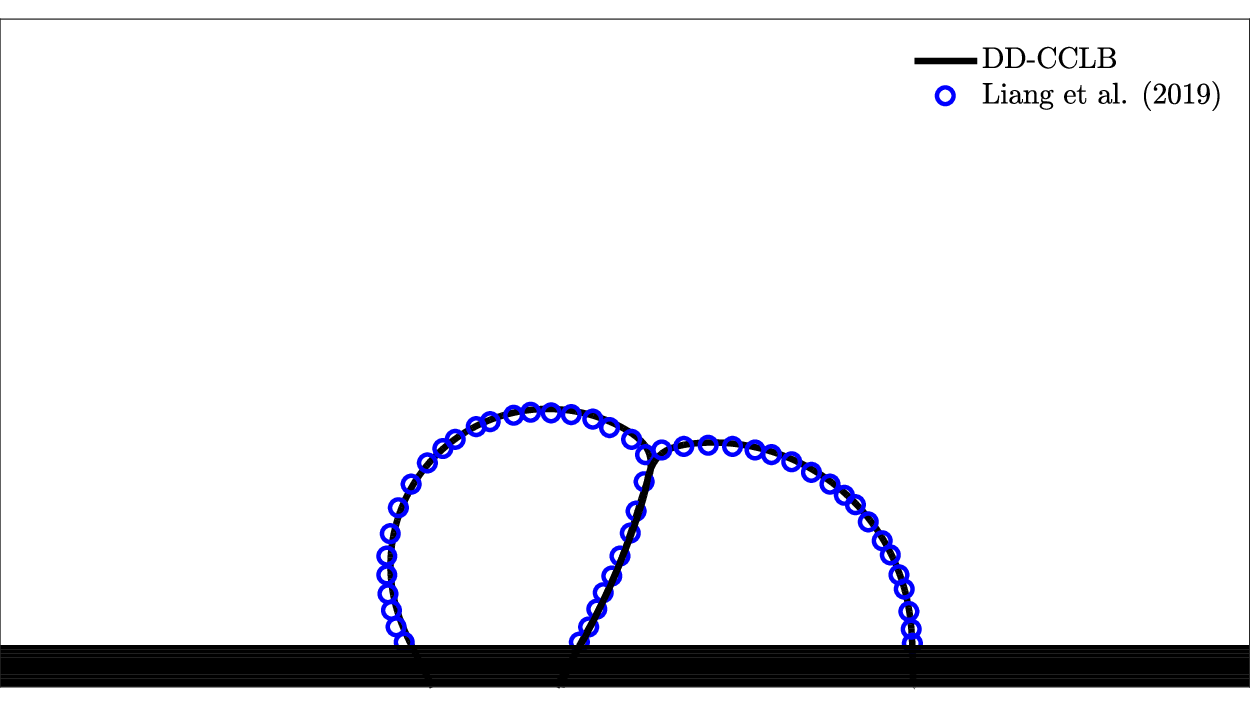}}
		\caption{The predicted equilibrium shape of a compound droplet in ternary system at a fixed contact angle $\theta_{23}=90^\circ$ [The solid line and circle  ($C=0.5$) are obtained by the present DD-CCLB method ($M=N=3$) and ternary-phase method \cite{Liang2019wet} under conditions of prescribed contact angles: (a) $\theta_{13}=60^\circ$, (b) $\theta_{13}=90^\circ$, (c) $\theta_{13}=120^\circ$].}
	\end{center}
	\label{fig:ex2}
\end{figure}

Now we investigate the spreading of a compound droplet ($M=N=3$) on an ideal wall to test the present DD-CCLB method for the ternary system. In our simulations, two immiscible fluids $C_1$ and $C_2$ with the same radius $R=60\delta x$ constitute a semicircular compound drop, which are surrounded by another fluid ($C_3$) and deposited on an ideal wall. The physical domain is divided into $300\delta x\times 170\delta x$, and some parameters are given by $\rho_1=100$, $\rho_2=50$, $\rho_3=1$, $\sigma_{ij}=0.005$ with $i\neq j$, $m_0=0.01$, $\varepsilon=8\delta x$, $\varepsilon_0=0.6$, $\delta x=0.1$, $c=10$, and other parameters are the same as those adopted in the previous problem.
We conduct some simulations, and present a comparison with the results in Ref.  \cite{Liang2019wet}, where the spreading of compound droplets on a substrate is studied by using the original governing equations combined the boundary condition (\ref{eq:2.7}) for ternary-phase flows. Figure \ref{fig:ex2} depicts the equilibrium shapes of a compound droplet at various contact angles with a fixed contact angle $\theta_{23}=90^{\circ}$, and the present numerical results are close to those reported in the previous work \cite{Liang2019wet}. In addition, to give a quantitative test on the accuracy of the present DD-CCLB method, we also measure the equilibrium spreading lengths of compound droplet, and list the corresponding analytical solutions and available numerical data in Table \ref{table2}, from which a good agreement between them can be observed.

\begin{table}
	\begin{small}
		\centering
		\caption{The equilibrium spreading lengths $L_1$ and $L_2$ (normalized by the initial radius $R$) in the spreading of a compound droplet in ternary system with $\theta_{23}=90^\circ$.}
		\begin{tabular}{lcccccccc}
			\toprule
			&&&& \multicolumn{2}{c}{Numerical} && \multicolumn{2}{c}{Relative errors}\\
			\cline{5-6}\cline{8-9}
			Case & Lengths & Analytical && \emph{DD-CCLB}, &Liang et al. \cite{Liang2019wet}&&  \emph{DD-CCLB}, & Liang et al. \cite{Liang2019wet}\\
			%		& & && $n=1$ & $n=2$ & && $n=1$ & $n=2$ & \\
			\midrule
			$\theta_{13}=60^\circ$ & $L_1$ & 1.707   &&   1.698   &1.683  && 0.53\% &1.41\% \\
			& $L_2$ & 1.072   &&    1.044  & 1.083  && 2.61\% &1.03\%\\                                                                                 	
			$\theta_{13}=90^\circ$  & $L_1$ & 1.183  &&   1.172   & 1.175  && 0.93\% &0.68\% \\
			& $L_2$ & 1.183  &&  1.172   & 1.192  && 0.93\% &0.76\% \\                                                                                     
			$\theta_{13}=120^\circ$ & $L_1$ & 0.672 && 0.677   & 0.688   &&0.74\% &2.38\% \\
			& $L_2$ & 1.335 &&  1.332    &1.337  && 0.22\% &0.15\% \\
			\bottomrule
		\end{tabular}
		\label{table2}
	\end{small}
\end{table}

\subsection{\label{sec:level4.3}The spreading of a compound droplet in quaternary system on an ideal wall}
\begin{figure}[H]
	\begin{center}
		\subfigure[]{
			\includegraphics[width=1.50in]{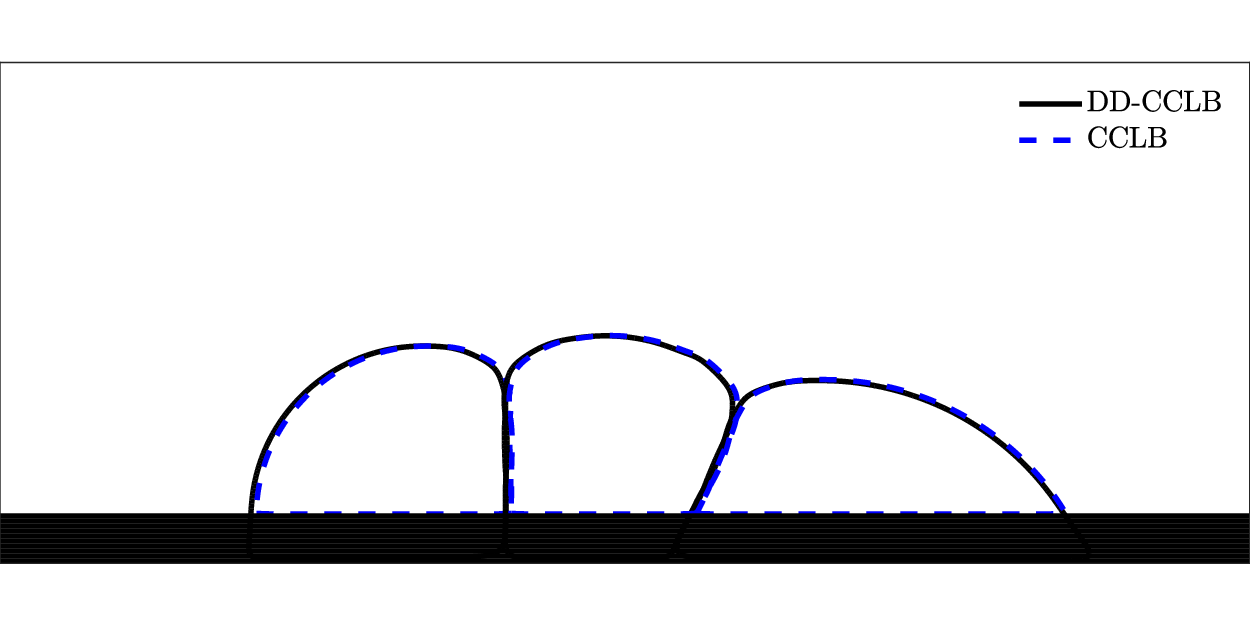}}
		\subfigure[]{ 
			\includegraphics[width=1.50in]{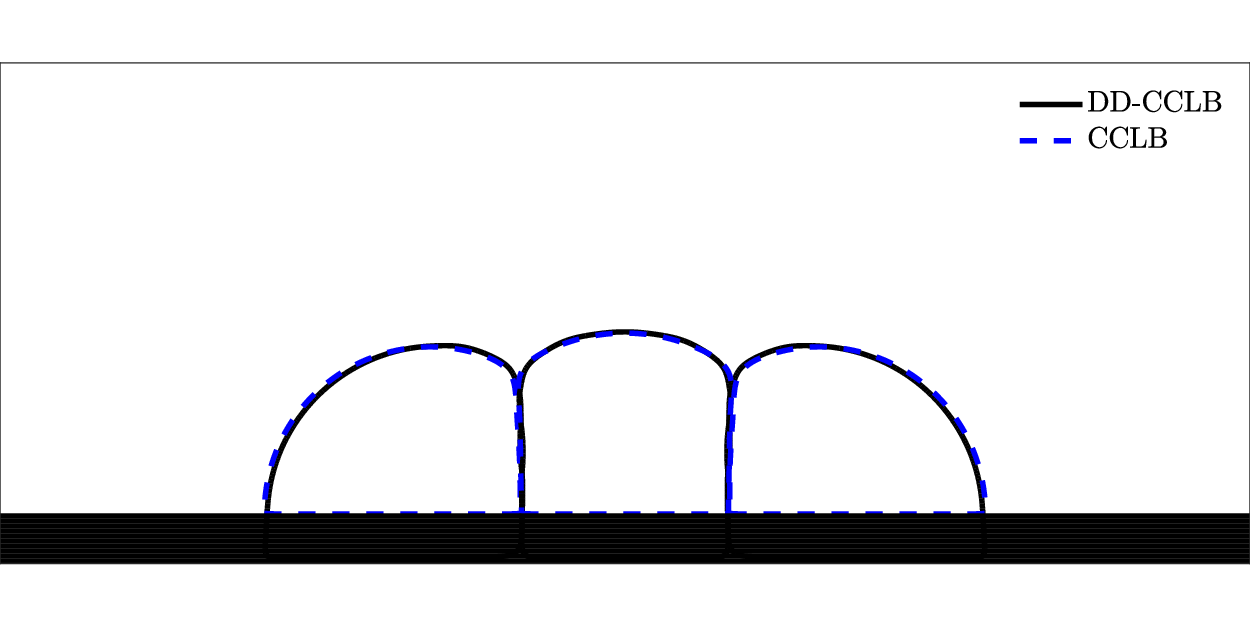}}
		\subfigure[]{ 
			\includegraphics[width=1.50in]{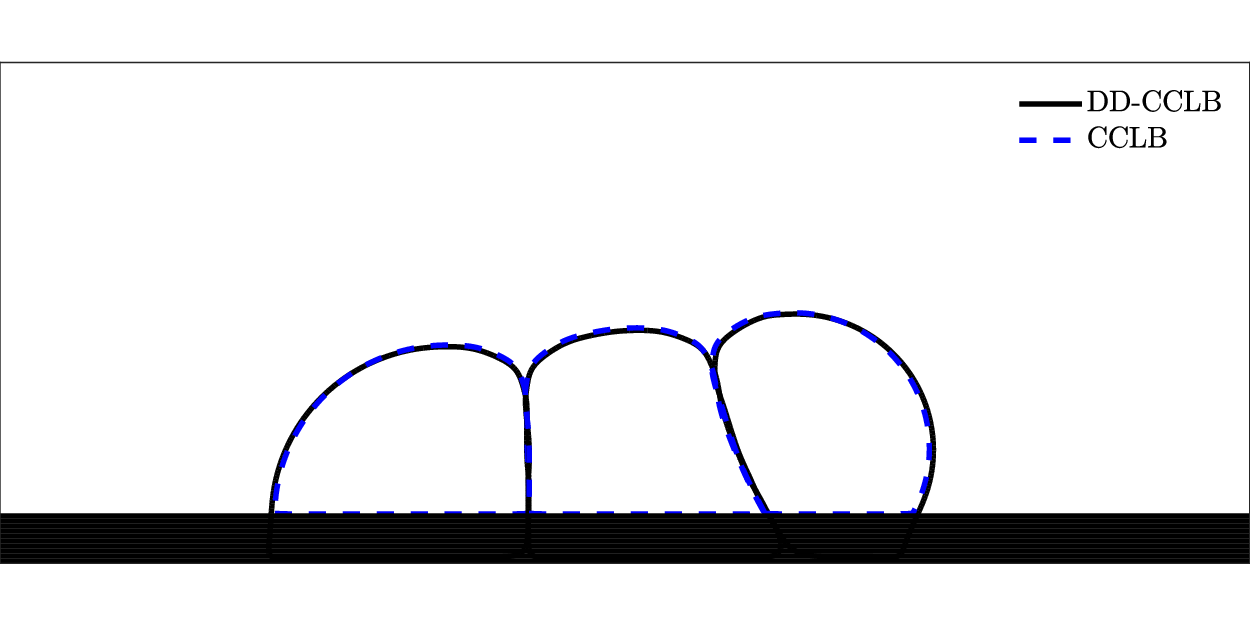}}
		\caption{The predicted equilibrium shape of a compound droplet in quaternary system at the fixed contact angles $\theta_{14}=\theta_{24}=90^\circ$ [The solid and dashed lines ($C=0.5$) are obtained by the present DD-CCLB method ($M=N=4$) and CCLB method under conditions of prescribed contact angles: (a) $\theta_{34}=60^\circ$, (b) $\theta_{34}=90^\circ$, (c) $\theta_{34}=120^\circ$].}
	\end{center}
	\label{fig:ex3}
\end{figure}

We continue to test the DD-CCLB method for multiphase flows through considering the spreading of a compound drop in quaternary system ($M=N=4$) on an ideal wall. In the physical domain $\Omega$ with the lattice size $500\delta x\times200\delta x$, three semicircular droplets with the same radius $R=60\delta x$ are contacted with each other, which are surrounded by the fourth phase and initially placed on an ideal wall. In the following simulations, the parameters are set as $\rho_1=1$, $\rho_2=0.8$, $\rho_3=0.5$, $\rho_4=0.1$, $\sigma_{ij}=0.005$ with $i\neq j$, $m_0=0.002$, $\varepsilon=10\delta x$, $\varepsilon_0=0.8$, $d_0=0.5$. We investigate the effect of the contact angle $\theta_{34}$ when $\theta_{14}=\theta_{24}=90^{\circ}$ are fixed, which leads to $\theta_{12}=90^{\circ}$ according to Eq. (\ref{eq:2.9}). From Fig. \ref{fig:ex3}, one can find that the equilibrium shapes of the quaternary-phase compound droplet are different by changing $\theta_{34}$. For example, when $\theta_{34}=60^{\circ}$, the third phase tends to spread on the wall with a larger spreading length $L_3$. With the increase of $\theta_{34}$, however, the third phase then begins to shrink on the substrate, and the length $L_3$ decreases. In particular, the compound droplet forms a symmetrical structure in case of $\theta_{34}=90^{\circ}$. We note that inspired by \cite{Zhang2016},
one can obtain the analytical solutions of the spreading lengths under the condition of $\theta_{12}=90^{\circ}$, and the details can be found in Appendix D. We also list the spreading lengths of the compound droplet with different contact angles in Table \ref{table3}, and give a comparison among the present numerical results, theoretical solutions, and those based on CCLB method which directly solves the original governing equations combined with the boundary condition (\ref{eq:2.7}). From Fig. \ref{fig:ex3} and Table \ref{table3}, one can find that the results obtained by different methods agree well with each other, which indicates that the DD-CCLB method is accurate in predicting the contact angle and equilibrium shape of a compound droplet on the ideal wall.

\begin{table}
	\centering
	\caption{The equilibrium spreading lengths $L_1$, $L_2$ and $L_3$ (normalized by the initial radius $R$) in the spreading of a compound droplet in quaternary system with $\theta_{14}=\theta_{24}=90^\circ$.}
	\begin{tabular}{lcccccccc}
		\toprule
		&&&& \multicolumn{2}{c}{Numerical} && \multicolumn{2}{c}{Relative errors}\\
		\cline{5-6}\cline{8-9}
		Case & Lengths & Analytical && \emph{DD-CCLB}, & \emph{CCLB}&&  \emph{DD-CCLB}, & \emph{CCLB}\\
		%		& & && $n=1$ & $n=2$ & && $n=1$ & $n=2$ & \\
		\midrule
		$\theta_{34}=60^\circ$& $L_1$ & 1.672 && 1.698 & 1.697    && 1.56\% & 1.50\% \\
		& $L_2$ & 1.273 && 1.247 & 1.238     && 2.04\% &2.75\%\\ 
		& $L_3$ & 2.416&& 2.472 & 2.458     &&2.32\% & 1.74\%\\                                                                     	
		$\theta_{34}=90^\circ$ & $L_1$ & 1.672&& 1.698 & 1.708     && 1.56\% & 2.15\%  \\
		& $L_2$ & 1.434 && 1.372 & 1.390    && 4.32\% & 3.07\% \\
		& $L_3$ & 1.672 && 1.698 & 1.707    && 1.56\% & 2.09\%  \\	                                                                                     
		$\theta_{34}=120^\circ$ & $L_1$ &1.672 && 1.710& 1.695       && 2.27\% &1.38\% \\
		& $L_2$ & 1.652&& 1.598 & 1.570     && 3.27\% &4.96\% \\
		& $L_3$ & 0.948&& 1.005& 0.993      && 6.01\% &4.75\% \\	                                                                                    	
		\bottomrule
	\end{tabular}
	\label{table3}
\end{table}

\subsection{\label{sec:level4.4}A compound droplet spreading on a solid sphere}
\begin{figure}[H]{
		\begin{center}
			\subfigure[]{ \label{fig:ex41}
				\includegraphics[width=1.5in]{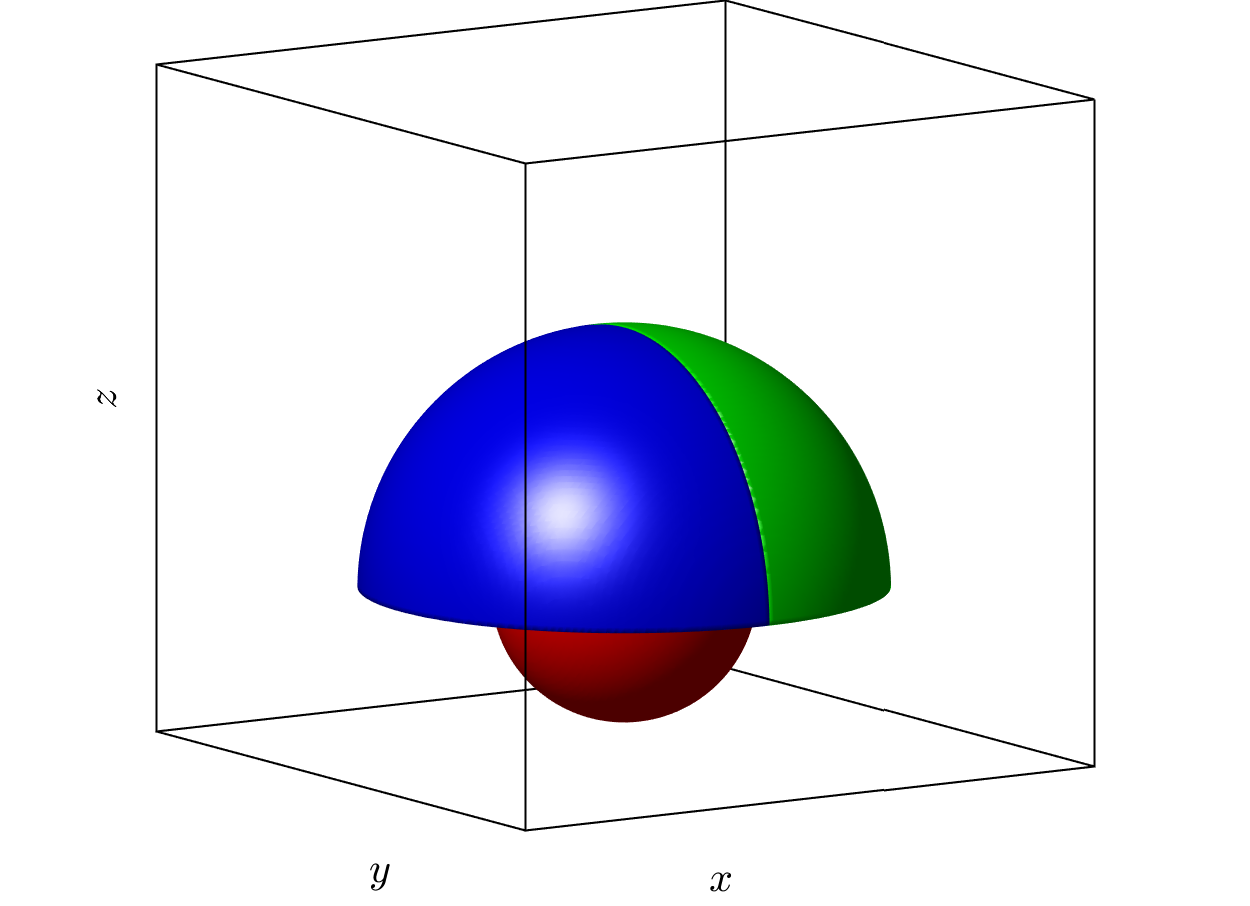}}
			\subfigure[]{ \label{fig:ex42}
				\includegraphics[width=1.0in]{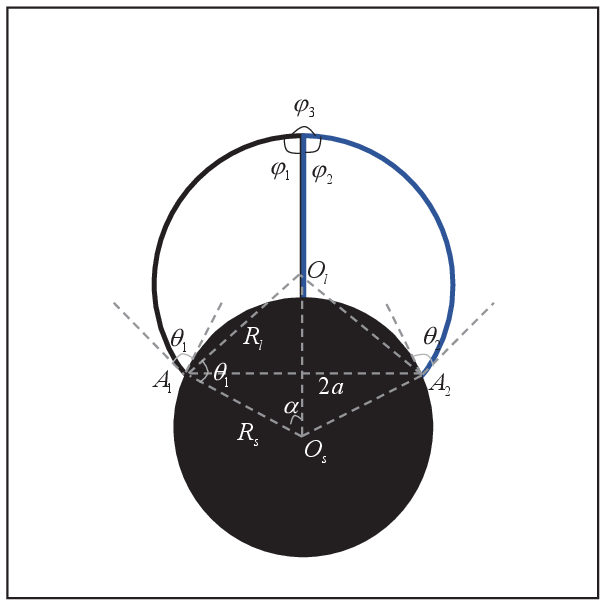}}
			\subfigure[]{ \label{fig:ex43}
				\includegraphics[width=1.0in]{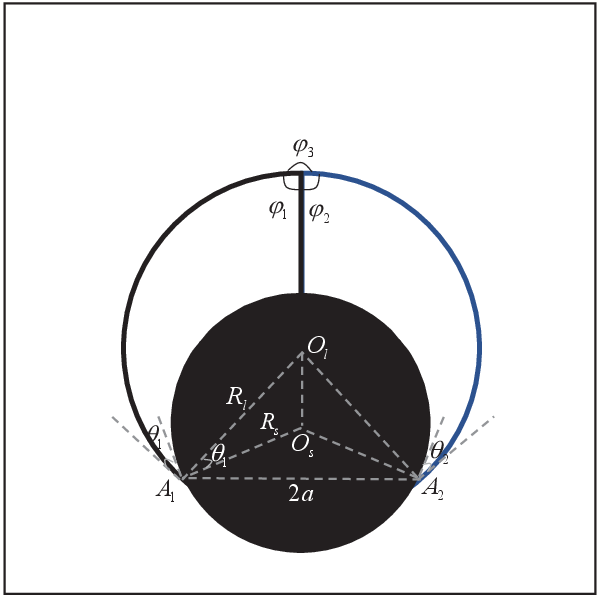}}
			\caption{(a) The initial setup for a hemispherical droplet on a solid sphere, (b) the profile in $z=0$ direction for a perfect compound droplet when $V_b<{V_s}/{2}$, (c) the profile in $z=0$ direction for a perfect compound droplet when $V_b>{V_s}/{2}$.}
	\end{center}}
\end{figure}
\begin{figure}[H]{
		\begin{center}
			\subfigure[]{
				\includegraphics[width=1.20in]{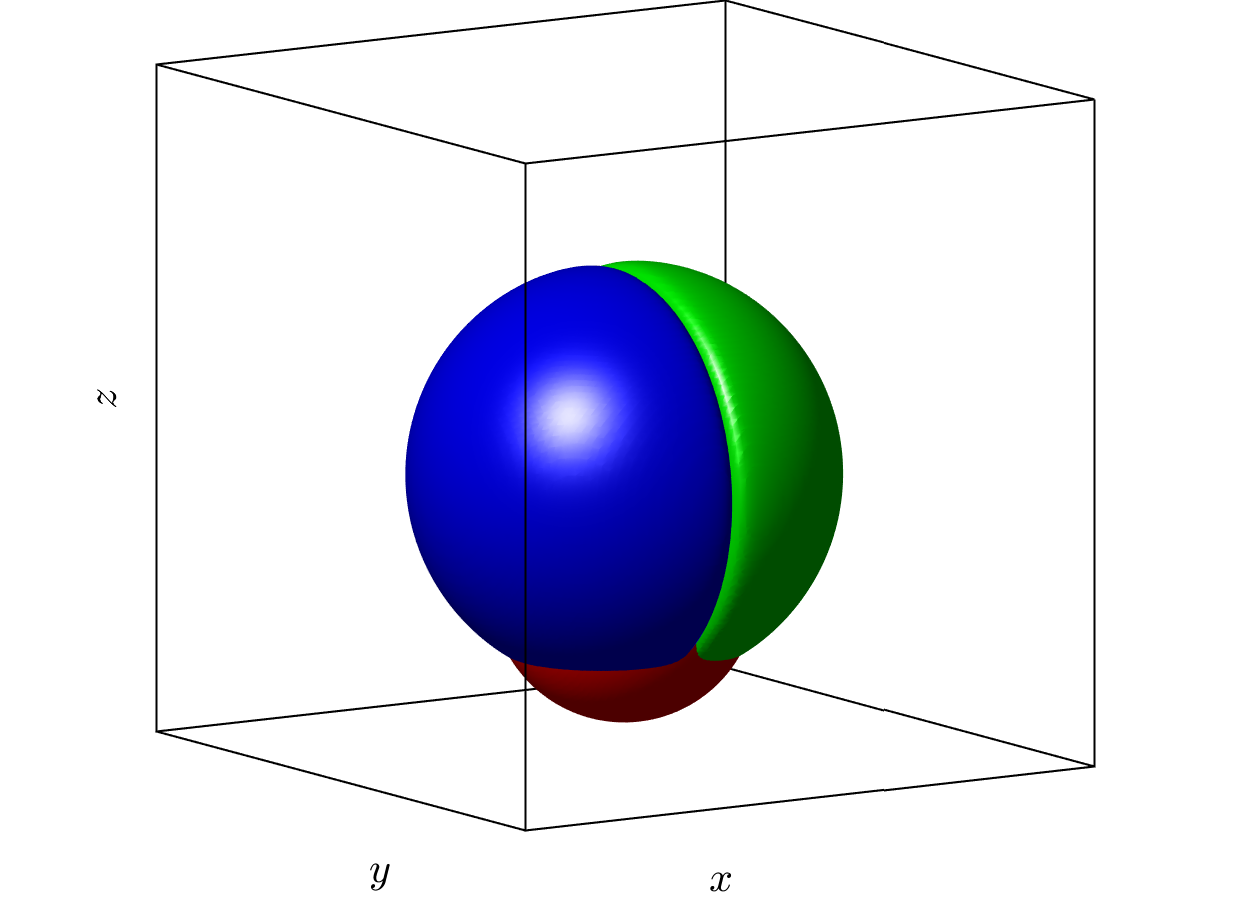}}
			\subfigure[]{
				\includegraphics[width=1.20in]{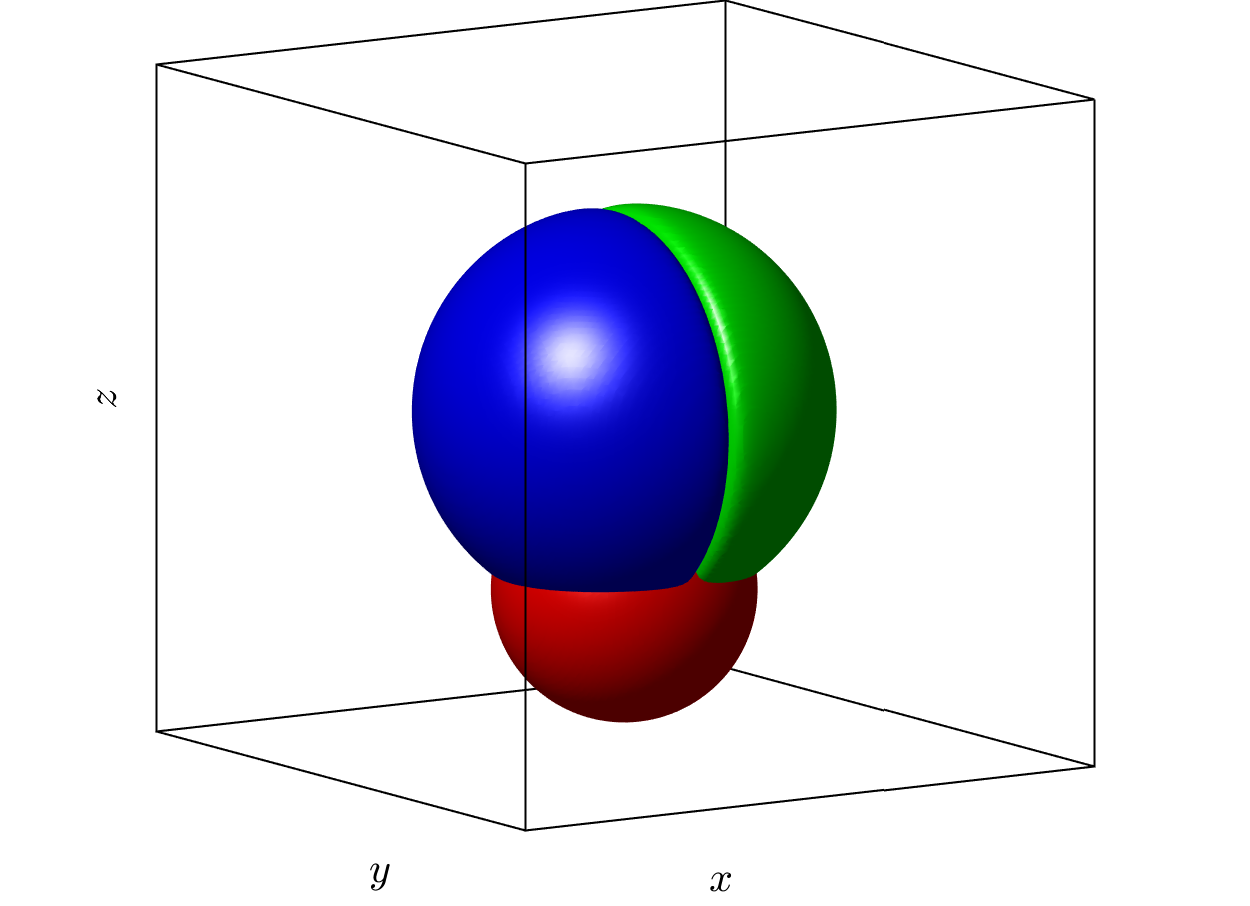}}
			\subfigure[]{ 
				\includegraphics[width=1.20in]{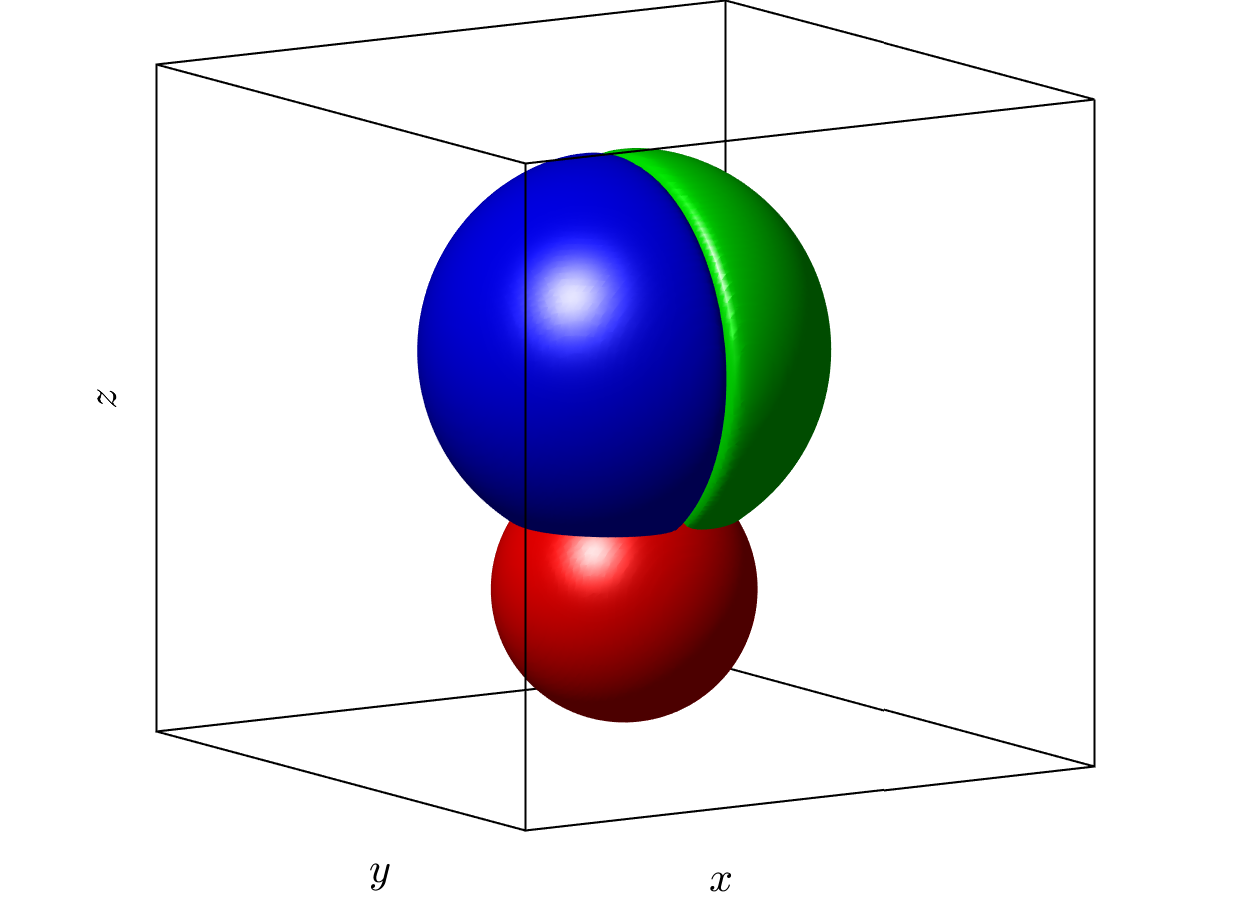}}
			\subfigure[]{ 
				\includegraphics[width=1.20in]{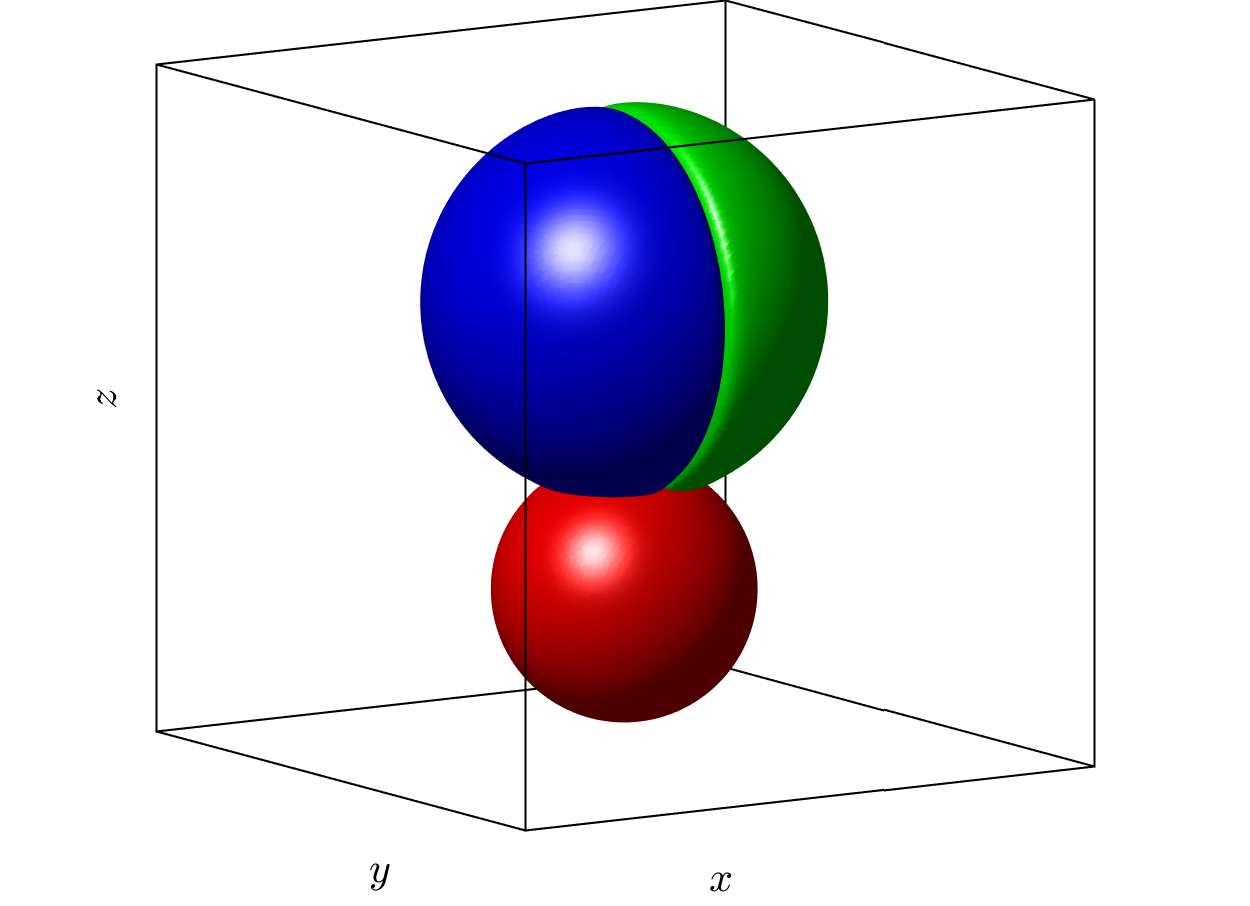}}		
			\subfigure[]{ 
				\includegraphics[width=1.20in]{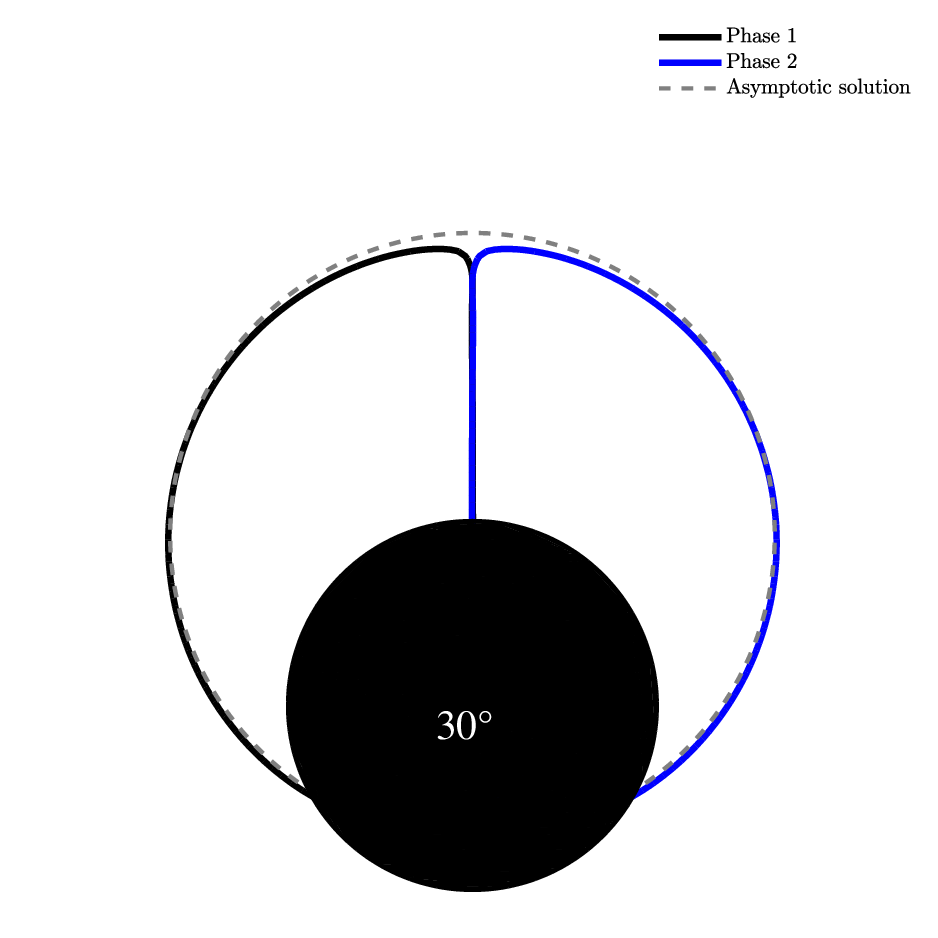}}
			\subfigure[]{ 
				\includegraphics[width=1.20in]{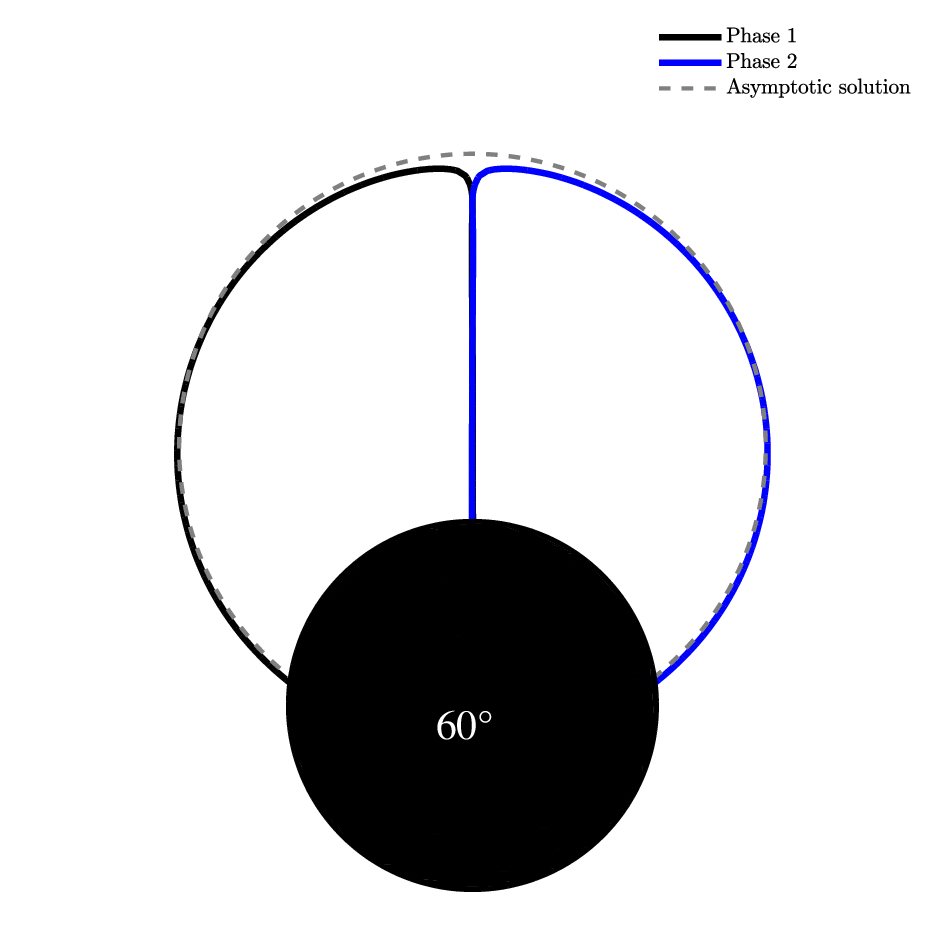}}
			\subfigure[]{
				\includegraphics[width=1.20in]{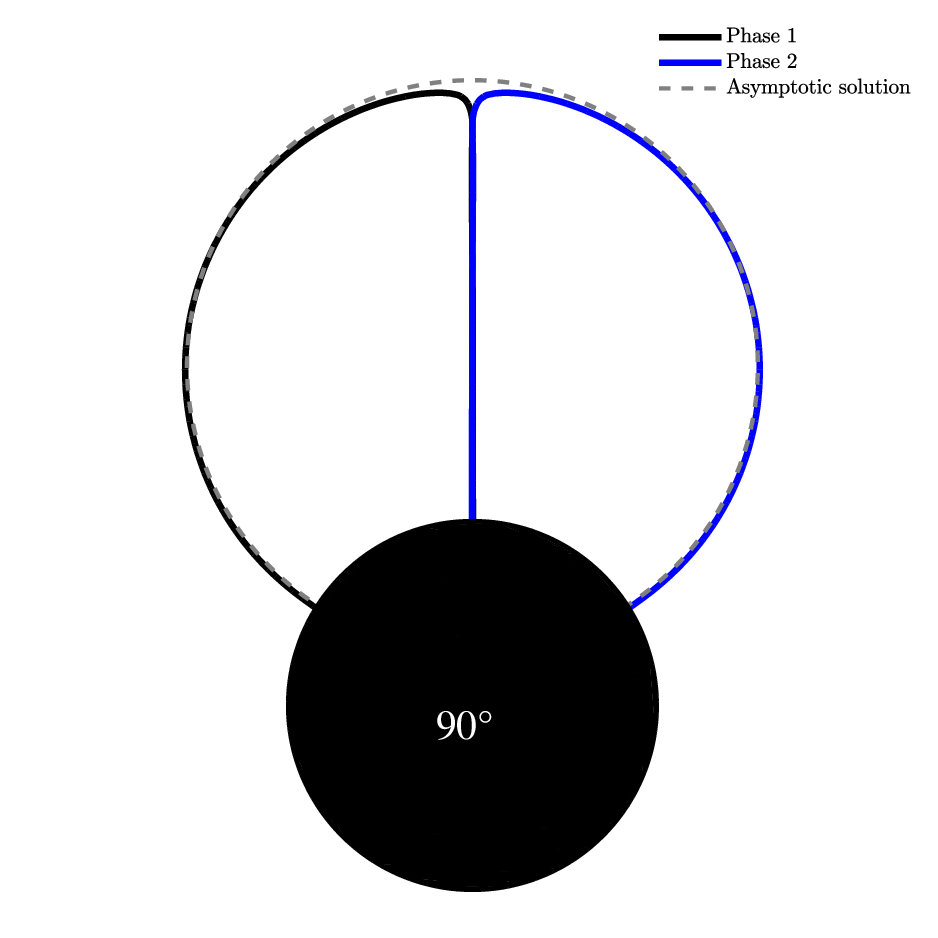}}
			\subfigure[]{
				\includegraphics[width=1.20in]{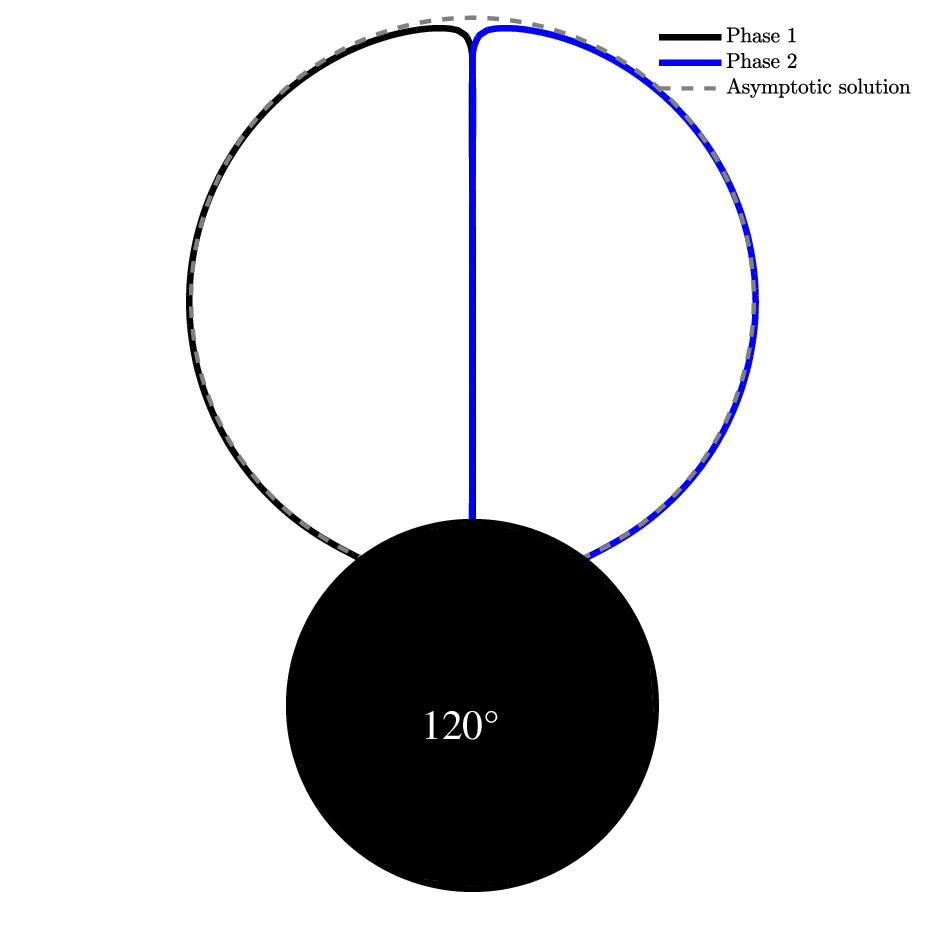}}		
			\caption{The predicted equilibrium shape of a compound droplet spreading on a solid sphere [The solid and dashed lines ($C=0.5$) in (e)-(h) are obtained by the present DD-CCLB method and the asymptotic solution under conditions of prescribed contact angles: (a) and (e) $\theta_{1,2}=30^\circ$, (b) and (f) $\theta_{1,2}=60^\circ$, (c) and (g) $\theta_{1,2}=90^\circ$, (d) and (h) $\theta_{1,2}=120^\circ$].}
	\end{center}}
	\label{fig:ex4}
\end{figure}
\begin{figure}[H]{
		\centering
		\includegraphics[width=2in]{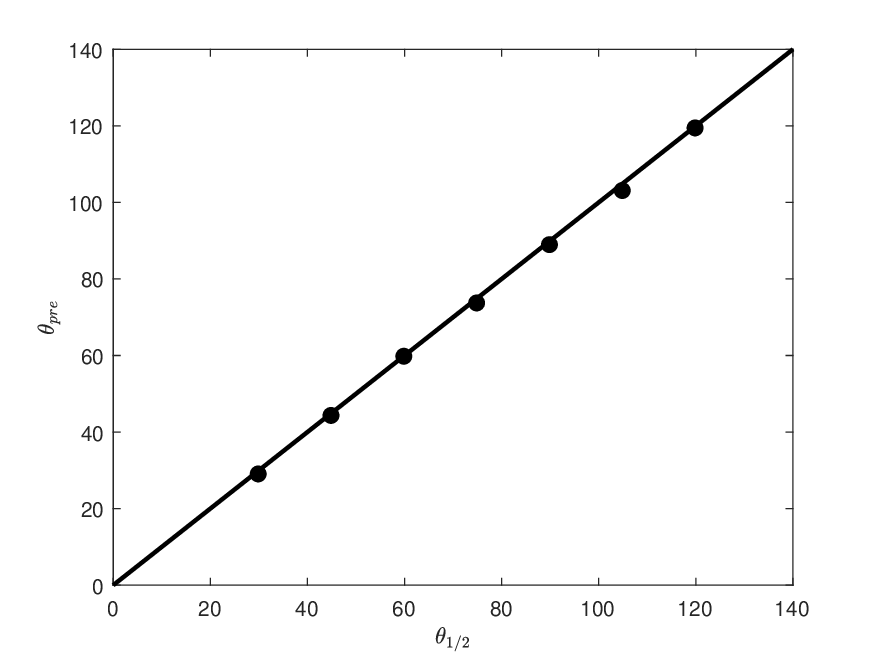}
		\caption{A comparison of the predicted contact angle ($\theta_{pre}$) and the specified contact angle ($\theta_{1,2}$).}}
	\label{Exp4.2:Fig3}
\end{figure}

The last benchmark problem we consider is a compound droplet spreading on a solid sphere, which is more complicated and can also be used to test the present LB method in predicting the contact angle of multiphase flow in a complex geometry. The schematic of the problem is shown in Fig. \ref{fig:ex41}, where a hemispherical compound droplet composed of blue ($C_1$) and green ($C_2$) fluids is initially placed on a solid sphere, and is immersed in another fluid ($C_3$). Under the assumption of two fluids ($C_1$ and $C_2$) with the same size, density, viscosity and surface wettability, the compound droplet would eventually form a spherical shape under the action of interfacial tensions, which can be regarded as a single-phase droplet. In this case, we can obtain the asymptotic solution of the equilibrium shape. As seen from Figs. \ref{fig:ex42} and \ref{fig:ex43}, when the droplet reaches a steady state, the contact angle $\theta_{pre}$ on spherical convex surface can be given by \cite{Extrand2012}
\begin{equation}
	\theta_{pre}=  2 \arctan \left\{\frac{\left[\frac{48 V_t}{\pi(2 a)^3}+\left(4+\left(\frac{48 V_t}{\pi(2 a)^3}\right)^2\right)^{\frac{1}{3}}\right]^{\frac{2}{3}}-2^{\frac{2}{3}}}{2^{\frac{1}{3}}\left[\frac{48 V_t}{\pi(2 a)^3}+\left(4+\left(\frac{48 V_t}{\pi(2 a)^3}\right)^2\right)^{\frac{1}{2}}\right]^{\frac{1}{3}}}\right\}
	-\arcsin \left(\frac{ 2a}{2R_s}\right),
	\label{eq:4.1}
\end{equation}
where $V_t=V_0+V_b$ with $V_0$ and $V_b$ being the volumes of the compound droplet and the spherical solid covered by the droplet. Actually, it is easy to obtain the initial volume $V_0$ as
\begin{equation}
	V_0=\frac{2\pi}{3 } R_0^3-\frac{2\pi}{3 } R_s^3,
\end{equation}
where $R_0$ is the initial radius of the hemispherical droplet and $R_s$ is the radius of the solid sphere. If $V_b$ is less than the half volume of the solid sphere $V_s$, see Fig. \ref{fig:ex42}, one can derive its expression as 
\begin{equation}
	V_b=\frac{1}{3} \pi R_s^3\left\{2-3\left[1-\left(\frac{2 a}{2 R_s}\right)^2\right]^{\frac{1}{2}}+\left[1-\left(\frac{2 a}{2 R_s}\right)^2\right]^{\frac{3}{2}}\right\}.
\end{equation}
On the contrary, if $V_b$ is larger than the half volume of the solid sphere $V_s$, see Fig. \ref{fig:ex43}, it can be determined by
\begin{equation}
	V_b=\frac{4}{3} \pi R_s^3-\frac{1}{3} \pi R_s^3\left\{2-3\left[1-\left(\frac{2 a}{2 R_s}\right)^2\right]^{\frac{1}{2}}+\left[1-\left(\frac{2 a}{2 R_s}\right)^2\right]^{\frac{3}{2}}\right\}.
\end{equation}
On the other hand, when the droplet is in equilibrium state with $R_l$, the volume of the droplet can also be given by
\begin{equation}
	V_0=\frac{4}{3} \pi R_l^3-\left[\frac{ \pi}{3} \left(3 R_s-h_1\right) h_1^2+\frac{ \pi}{3} \left(3 R_l-h_2\right) h_2^2\right],
\end{equation}
where $h_1=R_s(1-\cos\alpha)$ and $h_2=R_l[1-\cos(\pi-\theta_{pre}-\alpha)]$. 
Then one can determine the center of the steady droplet through calculating the distance $d_{O_jO_s}$, 
\begin{equation}
	d_{O_jO_s}=R_l^2+R_s^2-2 R_l R_s \cos\theta.
\end{equation}
From above analysis, we can predict the equilibrium shape of the compound droplet with the asymptotic solution. However, for a ternary-phase flow, there exits a triple point where the interfacial angles are $\varphi_1$, $\varphi_2$ and $\varphi_3$, and satisfy the condition $\varphi_1+\varphi_2+\varphi_3=2\pi$. Based on the balance of interfacial angles at the equilibrium state, one can obtain
\begin{equation}
	\frac{\sin \varphi_1}{\sigma_{23}}=\frac{\sin \varphi_2}{\sigma_{13}}=\frac{\sin \varphi_3}{\sigma_{12}} .
\end{equation}
Owing to the fact that  $\varphi_1=\varphi_2\approx\pi/2$ and $\varphi_3\approx\pi$, which can be seen from Figs. \ref{fig:ex42} and \ref{fig:ex43}, in the following simulations, we choose $\sigma_{12}=0.0025$ and $\sigma_{13}=\sigma_{23}=0.05$ to obtain $\varphi_1=\varphi_2=88.6^{\circ}$. Some other parameters are set as $\rho_1=\rho_2=10$, $\rho_3=1$, $R_0=50\delta x$, $R_s=25\delta x$, $m_0=0.01$, $\varepsilon_0=0.5$, $\varepsilon=5\delta x$, $\delta x=0.1$, $c=10$, $d_0=0.4$, $s^{f_i}_0=s^{f_i}_{21}=1.0$, $s^{f_i}_1=10/19$, $s^{g}_0=1.0$ and $s^{g}_1=s^{g}_3=s^{g}_4=1.2$. We perform some simulations, and present the numerical (solid curve) and asymptotic (dashed curve) solutions of the equilibrium droplet shape in Fig. \ref{fig:ex4} where the contact angles are $\theta_{1,2}=30^{\circ}$, $\theta_{1,2}=60^{\circ}$, $\theta_{1,2}=90^{\circ}$ and $\theta_{1,2}=120^{\circ}$. It is obvious that the numerical results agree well with the asymptotic solutions except for the difference at the triple point. Furthermore, Fig. \ref{Exp4.2:Fig3} shows a comparison between the predicted contact angle $\theta_{pre}$ based on Eq. (\ref{eq:4.1}) and the specified contact angle $\theta_{1,2}$. From this figure, one can find that there is a good agreement between the numerical results and the asymptotic solutions, which illustrates that the DD-CCLB method has a good performance in the study of the multiphase flow in complex solid surface.

\subsection{\label{sec:level4.5}A compound droplet passing through an irregular channel}
\begin{figure}[H]{
		\centering
		\includegraphics[width=2in]{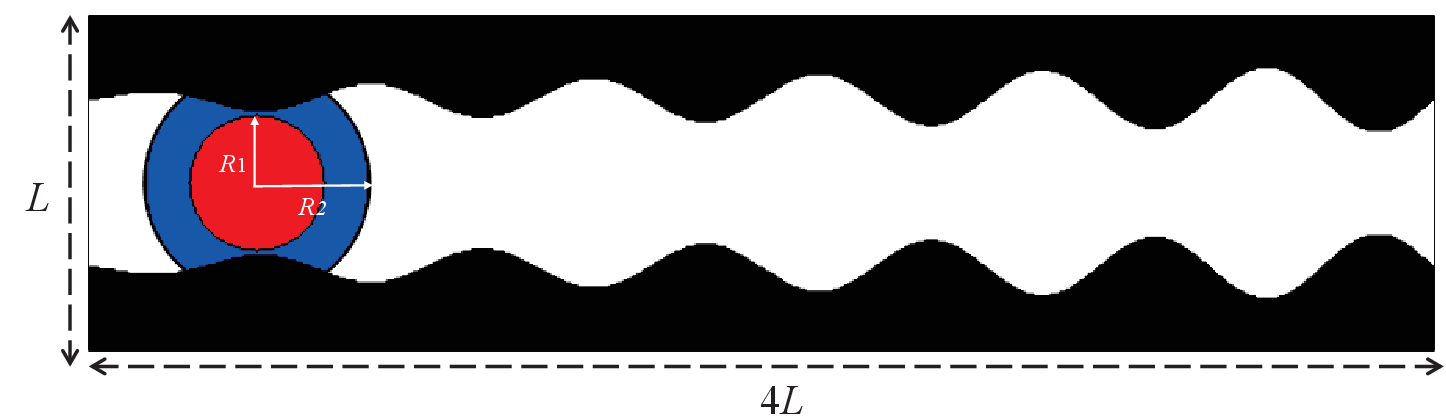}
		\caption{The schematic of a compound droplet passing through an irregular channel.}}
	\label{Exp5.0}
\end{figure}
\begin{figure}[H]{
		\begin{center}
			\subfigure[]{	\label{fig:ex5.60}
				\includegraphics[width=1.5in]{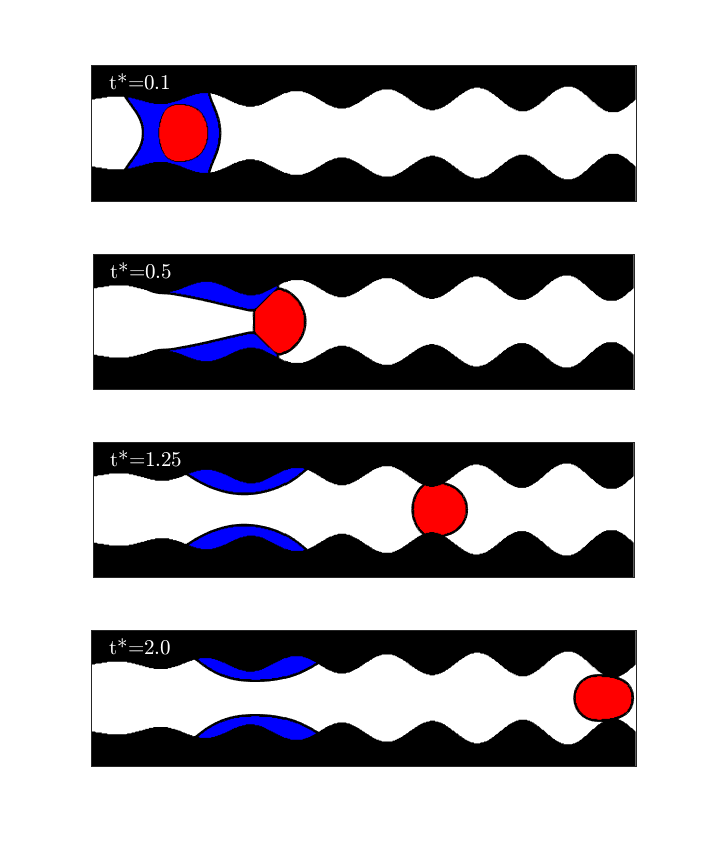}}
			\subfigure[]{   \label{fig:ex5.120}
				\includegraphics[width=1.5in]{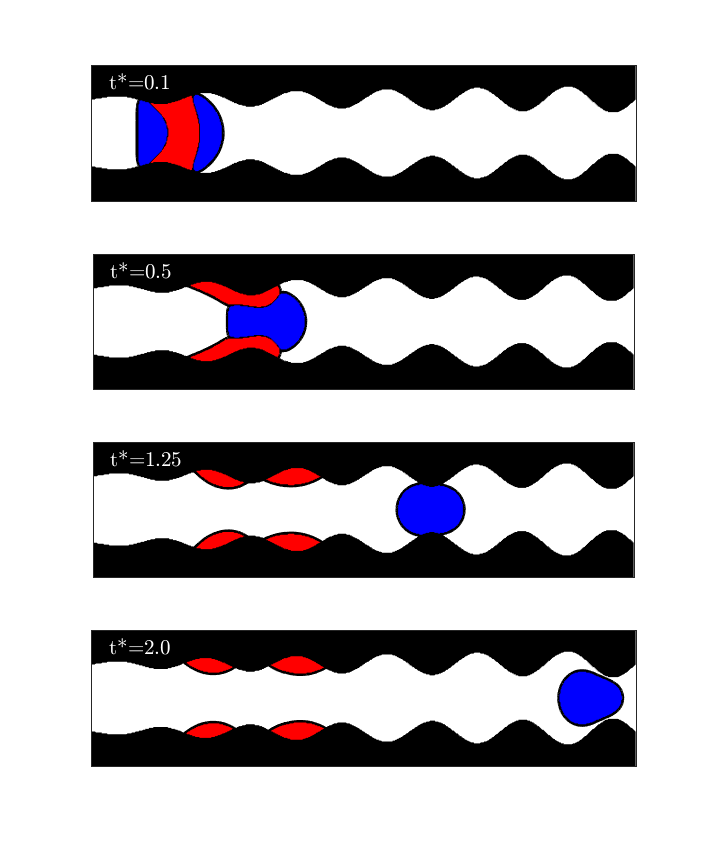}}
			\subfigure[]{ \label{fig:ex5.12060}
				\includegraphics[width=1.5in]{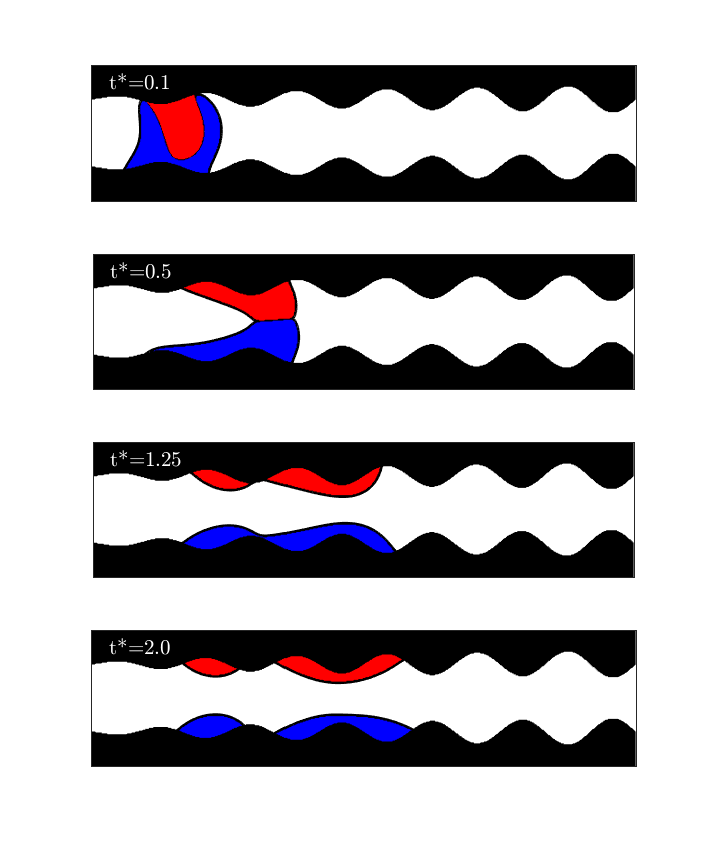}}		
			\caption{The dynamics of a compound droplet passing through an irregular channel with (a) $\theta^{1,2}_{13}=120^{\circ}$ and  $\theta^{1,2}_{23}=60^{\circ}$, (b) $\theta^{1,2}_{13}=60^{\circ}$ and  $\theta^{1,2}_{23}=120^{\circ}$, (c) $\theta^1_{13}=\theta^2_{23}=120^{\circ}$ and $\theta^{1}_{23}=\theta^{2}_{13}=60^{\circ}$ ( $t^*$ is normalized by $L/u_c$).}
	\end{center}}
	\label{fig:ex5}
\end{figure}

Unlike the static-state benchmark problems considered above, in this part, we investigate a compound droplet passing through the wavy channel, which is a typical irregular domain with a complex geometrical shape. The configuration of the problem is given in Fig. \ref{Exp5.0}, where a concentric droplet composed of the red inner droplet with $R_1=L/5$ and the blue outer droplet with $R_2=L/3$ is located at $(L/2, L/2)$ of the computational domain. Besides, the top and bottom boundaries of the wavy channel are defined by	$U(x)=-B(x)=0.05L\sqrt{x/L}\sin(3\pi x/L)$.
To reflect the effects of different wetting boundary conditions in the DD-CC equations, we introduce two characteristic functions,
\begin{equation}
	\phi_1(x, y)=\frac{1}{2}+\frac{1}{2} \tanh \left[\frac{2\left(y-L/4-B(x)\right)}{\varepsilon_0}\right],
\end{equation}
\begin{equation}
	\phi_2(x, y)=\frac{1}{2}-\frac{1}{2} \tanh \left[\frac{2\left(y-3L/4-U(x)\right)}{\varepsilon_0}\right],
\end{equation}
and then the characteristic function $\phi$ and the source term $\phi\mu_{C_i}$ appeared in the DD-CC equations can be rewritten as
\begin{equation}
	\phi(x, y)=	\phi_1(x, y)\phi_2(x, y),
\end{equation}
and 
\begin{equation}
	\begin{aligned}
		\phi\mu_{C_i}=\sum_{j=1}^N 2 \beta_{ij}\phi \left[g^{\prime}\left(C_i\right)-g^{\prime}\left(C_i+C_j\right)\right]+\sum_{j=1}^N	\frac{3\varepsilon}{4}\sigma_{ij} \nabla\cdot(\phi \nabla C_j)\\
		+\sum_{j=1}^N 3\sigma_{ij}\sum_{q\neq j}C_j C_q\left(\cos\theta^{1}_{jq}|\nabla\phi_1|+\cos\theta^2_{jq}|\nabla\phi_2|\right).
	\end{aligned}
\end{equation}
In our simulations, a parabolic velocity with the mean velocity $u_c = 0.003$ is imposed on the inlet and outlet of the channel, and the other parameters are fixed as $L=15$, $\rho_1=\rho_2=1.0$, $\rho_3=0.1$, $\sigma_{ij}=0.004$ where $i\neq j$, $m_0=0.01$, $\varepsilon_0=0.6$, $\varepsilon=8\delta x$, $\delta x=0.1$, $c=40$, $d_0=0.3$. We carry out some simulations, and plot the results in Fig. \ref{fig:ex5}. From this figure, one can see that the compound droplet can be separated in different patterns under different wetting properties of the solid walls. For example, as shown in Fig. \ref{fig:ex5.60} where $\theta^{1,2}_{13}=120^{\circ}$ and  $\theta^{1,2}_{23}=60^{\circ}$, the concentric droplet splits into two blue subdroplets adhered to the solid walls and an unbroken red droplet passes through the channel. In contrast, if $\theta^{1,2}_{13}=60^{\circ}$ and  $\theta^{1,2}_{23}=120^{\circ}$ [see Fig. \ref{fig:ex5.120}], the red subdroplets are adhered to the solid walls and the blue droplet can pass through the channel. For the case of $\theta^1_{13}=\theta^2_{23}=120^{\circ}$ and $\theta^{1}_{23}=\theta^{2}_{13}=60^{\circ}$ shown in Fig. \ref{fig:ex5.12060}, the red and blue droplets can be split from the concentric droplet and then adhered to the top and bottom walls, respectively. 
\section{\label{sec:level5}Conclusions}
In this paper, we first developed a DD  based conservative and consistent  phase-field model for the multiphase flows in complex geometries, which can handle the topological changes of the fluid-fluid interface as well as the irregular fluid-solid interface with different wettability properties. Then we proposed a new LB method for the DD-CC equations and tested the method by four benchmark problems. It is found that the present results are in good agreement with analytical solutions and/or numerical results based on other numerical methods, which illustrates that the DD-CCLB method has a good ability in capturing the multiphase interfacial changes and treating the complex boundary on the irregular solid surface. Finally, the present LB method is also used to study a compound droplet passing through an irregular channel, and the results show that the compound droplet can be separated in different forms under different wetting properties of the solid walls. In a future work, we will consider the DD-CCLB method for the multiphase flows in porous media, which is a more complicated problem and also important in many different fields (e.g., enhanced oil recovery).

\section*{Acknowledgements}
The computation is completed in the HPC Platform of Huazhong University of Science and Technology. This work is financially supported by the National Natural Science Foundation of China (Grants No. 12072127 and No. 51836003) and the Fundamental Research Funds for the Central Universities, HUST (No. 2021JYCXJJ010).
\appendix		
\section{Appendix: The Chapman-Enskog analysis of the LB model for the hybrid AC equation}
\appendix
\section{The matched asymptotic analysis on the DD-CH equations}\label{sec:appendixA}
\setcounter{equation}{0}
\renewcommand\theequation{A.\arabic{equation}}
Now we perform a matched asymptotic analysis to show that the DD-CH equation (\ref{eq:2.13a}) could converge to the CH equation (\ref{eq:2.4}) with the boundary conditions (\ref{eq:2.7}) as the interface width parameter $\varepsilon$ tends to zero. First, the DD variable $Y$, which represents $\textbf{u}$, $C_i$ or ${\mu}_{C_{i}}$, can be expanded in the powers of $\varepsilon$ in regions close to the boundary (inner expansion) and far from the boundary (outer expansion). On each side far from the domain boundary $\partial\tilde{\Omega}$, there exists an outer expansion for the variable $\bar{Y}_{i}$,
\begin{equation}
	\bar{Y}_{i}(\boldsymbol{x};\varepsilon)=\bar{Y}_{i}^{(0)}(\boldsymbol{x})+\varepsilon \bar{Y}_{i}^{(1)}(\boldsymbol{x})+\cdots, \quad i=1,2,
	\label{eq:A1}
\end{equation}
where $i=1$ denotes the inside $\tilde{\Omega}$ with $\phi=1$, and $i=2$ represents the outside $\tilde{\Omega}$ with $\phi=0$. Under the assumption of the boundary conditions on a larger and regular domain $\Omega$, $\textbf{n}\cdot\nabla \bar{\textbf{u}}_{2}=0$ and $\textbf{n}\cdot \nabla\bar{C}_{i2}=\textbf{n}\cdot\nabla\bar{\mu}_{{C}_{i2}}=0$, we can show that $	\bar{Y}_{2}$ satisfies the original equations automatically. 
For the case of $i=1$, inserting the expansion (\ref{eq:A1}) into Eqs. (\ref{eq:2.13a}) and (\ref{eq:2.14}), one can get
\begin{subequations}
	\begin{equation}
		\partial_{t}\bar{C}_{i1}^{(0)}+\nabla\cdot\left(\bar{C}_{i1}^{(0)}\bar{\textbf{u}}_{1}^{(0)}\right)=\sum_{ j=1}^N\nabla\cdot\left[M\left(\bar{C}_{i1}^{(0)},\bar{C}_{j1}^{(0)}\right)\nabla\bar{\mu}_{C_{i1}}^{(0)}\right],
	\end{equation}
	\begin{equation}
		\bar{\mu}_{C_{i1}}^{(0)}=\sum_{j=1}^N 2 \beta_{ij}\left[g^{\prime}\left(	\bar{C}_{i1}^{(0)}\right)-g^{\prime}\left(\bar{C}_{i1}^{(0)}+\bar{C}_{j1}^{(0)}\right)\right]+\sum_{j=1}^N	\frac{3\varepsilon}{4}\sigma_{ij} \nabla\cdot \nabla \bar{C}_{j1}^{(0)}.
	\end{equation}
\end{subequations}
If $\bar{Y}^{0}_{1}$ satisfies the corresponding boundary condition (\ref{eq:2.7}) on $\partial \tilde{\Omega}$, $\bar{Y}_{1}$ is the unique solution of CH equation, which means that the CH equation can be recovered by the DD-CH equation at the leading order $\mathcal{O}\left(\varepsilon\right)$. 

Then, we conduct the inner expansion for $\hat{Y}$ associated with local coordinate variables $z$ and $\textbf{s}$ as
\begin{equation}
	\hat{Y}(z, \textbf{s} ; \varepsilon)=\hat{Y}^{(0)}(z, \textbf{s})+\varepsilon \hat{Y}^{(1)}(z, \textbf{s})+\cdots,
	\label{eq:A9}
\end{equation}
where $\phi=1$ as ${z \rightarrow-\infty}$, and $\phi=0$ as ${z \rightarrow+\infty}$. In a overlapped region where both expansions are valid, the following asymptotic matching conditions at the first two leading orders can be derived,
\begin{equation}
	\begin{aligned}
		\lim _{z \rightarrow-\infty} \hat{Y}^{(0)}(z, \textbf{s})=\bar{Y}_{1}^{(0)}(\textbf{s}), \quad 	\lim _{z \rightarrow-\infty} \hat{Y}^{(1)}(z, \textbf{s})=\bar{Y}_{1}^{(1)}(\textbf{s})+z \textbf{n} \cdot \nabla \bar{Y}_{1}^{(0)}, \\
		\lim _{z \rightarrow+\infty} \hat{Y}^{(0)}(z, \textbf{s})=\bar{Y}_{2}^{(0)}(\textbf{s}), 
		\quad \lim _{z \rightarrow+\infty} \hat{Y}^{(1)}(z, \textbf{s})=\bar{Y}_{2}^{(1)}(\textbf{s})+z \textbf{n} \cdot \nabla \bar{Y}_{2}^{(0)}.
	\end{aligned}
	\label{eq:A4}
\end{equation}
At the leading order $\mathcal{O}\left(\varepsilon^{-2}\right)$, one can obtain
\begin{equation}
	\text {Eq. (\ref{eq:2.13a})} \rightarrow \sum_{ j=1}^N\left[\phi M\left(\hat{C}_{i}^{(0)},\hat{C}_{j}^{(0)}\right)\hat{\mu}^{(0)}_{C_{jz}}\right]_{z}=0, 
\end{equation}
\begin{equation}
	\text {Eq. (\ref{eq:2.14})} \rightarrow
	\sum_{j=1}^N	\frac{3\varepsilon}{4}\sigma_{ij}
	\left(\phi\hat{C}_{jz}^{(0)}\right)_{z}=0 .
\end{equation}
At the next order $\mathcal{O}\left(\varepsilon^{-1}\right)$, we have
\begin{equation}
	\text {Eq. (\ref{eq:2.13a})} \rightarrow \sum_{ j=1}^N\left[\phi M\left(\hat{C}_{i}^{(0)},\hat{C}_{j}^{(0)}\right)\hat{\mu}^{(1)}_{C_{jz}}\right]_{z}=0, 
\end{equation}
\begin{equation}
	\text {Eq. (\ref{eq:2.14})} \rightarrow \sum_{j=1}^N	\frac{3\varepsilon}{4}\sigma_{ij}
	\left(\phi\hat{C}_{jz}^{(1)}\right)_{z}=-\sum_{j=1}^N 3\sigma_{ij}\phi_z\sum_{q\neq j}\cos\theta_{jq}\bar{C}_j^{(0)} \bar{C}_q^{(0)}.
\end{equation}
Integrating above equations from $-\infty$ to $+\infty$ with respect to $z$, one can obtain
\begin{equation}
	\lim _{z \rightarrow-\infty}\hat{\mu}^{(1)}_{C_{iz}}=0, \quad
	\lim _{z \rightarrow-\infty} \hat{C}_{iz}^{(1)}=\sum_{j\neq i}\frac{4}{\varepsilon}\cos\theta_{ij}\bar{C}_i^{(0)} \bar{C}_j^{(0)}.
\end{equation}
Using the matching condition (\ref{eq:A4}), we can derive
\begin{equation}
	\textbf{n}\cdot \nabla\bar{\mu}_{C_{i1}}^{(0)}=0, \quad
	\textbf{n}\cdot \nabla \bar{C}_{i1}^{(0)}=\sum_{j\neq i}\frac{4}{\varepsilon}\cos\theta_{ij}\bar{C}_i^{(0)} \bar{C}_j^{(0)}.
\end{equation}
The outer solutions $\bar{\mu}_{C_{i1}}^{(0)}$ and $\bar{C}_{i1}^{(0)}$ satisfy the original boundary conditions (\ref{eq:2.7}), which indicates the DD-CH equation can recover the original CH equation at the leading order.

\section{The direct Taylor expansion of the DD-CCLB method for the DD-CC equations}\label{sec:appendixB}
%\appendix{\label{sec:appendix}}
\setcounter{equation}{0}
\renewcommand\theequation{B.\arabic{equation}}
We now carry out the direct Taylor expansion of Eqs. (\ref{eq:3.1}) and (\ref{eq:3.2}) to derive the DD-CC equations (\ref{eq:2.13}). First, we apply the Taylor expansion to the left hand sides of Eqs. (\ref{eq:3.1}) and (\ref{eq:3.2}) \cite{Chai2020Multiple,Chai2023Multiple,Chen1998LATTICEBM,Krueger2016TheLB,Succi2001TheLB},
\begin{equation}
	\sum_{l=1}^N \frac{\delta t^l}{l !} D_p^l f^i_p+O\left(\delta t^{N+1}\right)=-\Lambda_{p q}^{f^i}\left(f^i_q-f_q^{i,eq}\right)+\delta t\left(\delta_{pq}-\frac{\Lambda_{pq}^{f^i}}{2}\right) F^i_q,
	\label{eq:B1}
\end{equation}
\begin{equation}
	\sum_{l=1}^N \frac{\delta t^l}{l !} D_p^l g_p+O\left(\delta t^{N+1}\right)=-\Lambda_{p q}^g\left(g_q-g_q^{eq}\right)+\delta t\left(\delta_{pq}-\frac{\Lambda_{pq}^g}{2}\right) G_q,
	\label{eq:B2}
\end{equation}
where $D_p=\partial_t+\mathbf{c}_p \cdot \nabla$. Based on above equations, $f^i_p=f_p^{i,eq}+f_p^{i,ne}$ and $g_p=g_p^{eq}+g_p^{n e}$, we can find $f_p^{i,n e}=O(\delta t)$ and $g_p^{n e}=O(\delta t)$.
According to Eqs. (\ref{eq:B1}) and (\ref{eq:B2}), the following equations at different orders of $\delta t$ can be obtained,
\begin{equation}
	\begin{aligned}
		D_p f_p^{i,eq} =-\frac{\Lambda^{f^i}_{pq}}{\delta t} f_q^{i,n e}+\left(\delta_{pq}-\frac{\Lambda^{f^i}_{pq}}{2}\right) F^i_q+O(\delta t), \\
		D_p f_p^{i,eq}+D_p\left(\delta_{pq}-\frac{\Lambda^{f^i}_{pq}}{2}\right)\left(f_q^{i,n e}+\frac{\delta t}{2} F^i_q\right) =-\frac{\Lambda^{f^i}_{pq}}{\delta t} f_q^{i,n e}+\left(\delta_{pq}-\frac{\Lambda^{f^i}_{pq}}{2}\right) F^i_q\\
		+O\left(\delta t^2\right),
	\end{aligned}
	\label{eq:B3}
\end{equation}
\begin{equation}
	\begin{aligned}
		D_p g_p^{eq} & =-\frac{\Lambda^{g}_{pq}}{\delta t} g_q^{n e}+\left(\delta_{pq}-\frac{\Lambda^{g}_{pq}}{2}\right) G_q+O(\delta t), \\
		D_p g_p^{eq}+D_p\left(\delta_{pq}-\frac{\Lambda^{g}_{pq}}{2}\right)\left(g_q^{n e}+\frac{\delta t}{2} G_q\right) & =-\frac{\Lambda^{g}_{pq}}{\delta t} g_q^{n e}+\left(\delta_{pq}-\frac{\Lambda^{g}_{pq}}{2}\right) G_q+O\left(\delta t^2\right).
	\end{aligned}
	\label{eq:B4}
\end{equation}
From Eqs. (\ref{eq:3.5}), (\ref{eq:3.6}), (\ref{eq:3.8}) and (\ref{eq:3.10}), one can see that ${\Lambda}_{pq}$, $f_{p}^{i,eq}$, $F^i_p$, $g_{p}^{eq}$ and $G_{p}$ satisfy the following conditions,
\begin{subequations}
	\begin{equation}
		\sum \mathbf{e}_p {\Lambda}_{pq}  =s_0 \mathbf{e}_q,\quad
		\sum \mathbf{c}_p{\Lambda}_{pq}  =\mathbf{S}_1 \mathbf{c}_q,\quad
		\sum \mathbf{c}_p \mathbf{c}_p {\Lambda}_{pq} 
		=\mathbf{S}_{2}(\mathbf{c}_{q} \mathbf{c}_{q}-\frac{\mathbf{c}_{q} \cdot\mathbf{c}_{q}}{d}\mathbf{I})+c_s^2s_0 \mathbf{e}_q\mathbf{I},
		\label{eq:B.5a}
	\end{equation}
	\begin{equation}
		\sum_pf_{p}^{i,eq}=\phi C_i, \quad 
		\sum_p\textbf{c}_p f_{p}^{i,eq}=\phi C_i\textbf{u}, \quad
		\sum_p\textbf{c}_p \textbf{c}_p f_{p}^{i,eq}=c_s^2\eta_i\phi\mu_{C_i}\textbf{I},
		\label{eq:B.5b}
	\end{equation}
	\begin{equation}
		\sum_pF^{i}_{p}=0, \quad 
		\sum_p\textbf{c}_p F^{i}_{p}=\partial_t(\phi C_i\textbf{u})+c_s^2\eta_i(\mu_{C_i}\nabla\phi-\textbf{S}_i),
		\label{eq:B.5c}
	\end{equation}
	\begin{equation}
		\begin{aligned}
			\sum_pg_{p}^{eq}=\rho_0, \quad 
			\sum_p\textbf{c}_p g_{p}^{eq}=\phi\rho\textbf{u},\\
			\sum_p\textbf{c}_p \textbf{c}_p g_{p}^{eq}=\phi\rho\textbf{u}\textbf{u}+\frac{\phi\textbf{m}^{C}\textbf{u}+\phi\textbf{u}\textbf{m}^{C}}{2}+\phi p\textbf{I},\\
			\sum_p\textbf{c}_p \textbf{c}_p\textbf{c}_pg_{p}^{eq}=\phi\rho(c_s^2\mathbf{\Delta} +\bar{\delta}^{(4)})\cdot\textbf{u},
		\end{aligned}
		\label{eq:B.5d}
	\end{equation}
	\begin{equation}
		\sum_pG_{p}=\phi\textbf{u}\cdot\nabla\rho, \quad 
		\sum_p\textbf{c}_p G_{p}=\textbf{F},\quad 
		\sum_p\textbf{c}_p \textbf{c}_p G_{p}=c_s^2 \phi\mathbf{u} \cdot \nabla \rho \mathbf{I}+\mathbf{M}^{2G},
		\label{eq:B.5e}
	\end{equation}
	\label{eq:B.5}
\end{subequations}
where $s_0$ is the relaxation parameter, $\mathbf{S}_1$ and $\mathbf{S}_2$ are two $d\times d$ and $d^2\times d^2$ invertible relaxation matrices, $\mathbf{M}^{2G}$ is a second-order tensor to be determined below. $\Delta_{\alpha\beta\gamma\eta}=\delta_{\alpha\beta}\delta_{\gamma\eta}+\delta_{\alpha\gamma}\delta_{\beta\eta}+\delta_{\beta\gamma}\delta_{\alpha\eta}$ and $\bar{\delta}^{(4)}$ is defined by $\bar{\delta}^{(4)}=c^2-3c_s^2$ with $\alpha=\beta=\gamma=\eta$, otherwise $\bar{\delta}^{(4)}=0$. In addition, from Eqs. (\ref{eq:B4}), (\ref{eq:B.5d}) and (\ref{eq:B.5e}), we have
\begin{equation}
	\sum_pg_{p}^{ne}=-\frac{\delta_t}{2} \sum_pG_{p}=-\frac{\delta_t}{2}\phi\textbf{u}\cdot\nabla\rho, \quad 
	\sum_p\textbf{c}_{p}g_{p}^{ne}=-\frac{\delta_t}{2}	\sum_p\textbf{c}_{p}G_{p}=-\frac{\delta_t }{2}\textbf{F}.
\end{equation}
Summing Eq. (\ref{eq:B3}) over $p$, one can obtain
\begin{equation}
	\begin{aligned}
		\partial_{t} (\phi C_i) + \nabla \cdot (\phi C_i\textbf{u})&=O(\delta t),\\	
		\partial_{t} (\phi C_i) + \nabla \cdot (\phi C_i\textbf{u})+ \nabla \cdot\left(\textbf{I}-\frac{\textbf{S}^{{f}^i}_{1}}{2}\right) \left[ \sum_p\textbf{c}_pf_p^{i,ne}+\frac{\delta_t}{2}\sum_p\textbf{c}_pF_p^{i} \right]&=O(\delta t^2).	
	\end{aligned}
	\label{eq:B7}
\end{equation}
With the aid of Eq. (\ref{eq:B3}), we can determine $\sum_p\textbf{c}_pf_p^{i,ne}$ as
\begin{equation}
	\sum_p\textbf{c}_pf_p^{i,ne}=-\delta_t (\textbf{S}^{{f}^i}_{1})^{-1} \left[\partial_{t}(\phi C_i\textbf{u})+c_s^2\nabla\eta_i\phi\mu_{C_i}\textbf{I}-\left(\textbf{I}-\frac{{\textbf{S}}_1^{{f}^i}}{2}\right)\sum_p\textbf{c}_pF_p^{i}\right]+O(\delta t^2).
\end{equation}
Combining Eq. (\ref{eq:B7}) with $m_0=(1/s^{{f}^i}_{1}-1/2)c_s^2\eta_i\delta t$, yields Eq. (\ref{eq:2.13a}) at the order of $O(\delta t^2)$.
Following the same way, summing Eq. (\ref{eq:B4}) and $\textbf{c}_p$ $\times$ Eq. (\ref{eq:B4}), one can obtain
\begin{equation}
	\partial_{t}\phi\rho_0+	\nabla\cdot(\phi\rho\textbf{u})=\phi\textbf{u}\cdot\nabla\rho+O(\delta t^2),\\
\end{equation}
\begin{equation}
	\begin{aligned}
		\partial_{t}(\phi\rho\textbf{u})&+\nabla\cdot\left(\phi\rho\textbf{u}\textbf{u}+\frac{\phi\textbf{m}^{C}\textbf{u}+\phi\textbf{u}\textbf{m}^{C}}{2}+\phi p\textbf{I}\right)\\
		&+\nabla\cdot\sum_p \mathbf{c}_p \mathbf{c}_p\left(\delta_{pq}-\frac{\Lambda^{g}_{pq}}{2}\right)\left( g_p^{n e}+\frac{\delta t}{2}G_q\right)=\textbf{F}+O(\delta t^2).
	\end{aligned}
	\label{eq:B9}
\end{equation}
With the help of Eq. (\ref{eq:B4}), we can rewrite Eq. (\ref{eq:B9}) as
\begin{equation}
	\begin{aligned}
		\partial_{t}(\phi\rho\textbf{u})&+\nabla\cdot\left(\phi\rho\textbf{u}\textbf{u}+\frac{\phi\textbf{m}^{C}\textbf{u}+\phi\textbf{u}\textbf{m}^{C}}{2}+\phi p\textbf{I}\right)\\
		&-\nabla\cdot\delta_t\sum_p \mathbf{c}_p \mathbf{c}_p\left[\left(\Lambda^{g}_{pq}\right)^{-1}-\frac{\delta_{pq}}{2}\right]\left( 	D_q g_q^{eq}-G_q\right)=\textbf{F}+O(\delta t^2).
	\end{aligned}
	\label{eq:B11}
\end{equation}
In addition, from Eqs. (\ref{eq:B.5a}), (\ref{eq:B.5d}) and (\ref{eq:B.5e}), we get 	
\begin{equation}
	\begin{aligned}
		&\sum_p \mathbf{c}_p \mathbf{c}_p\left[\left(\Lambda^{g}_{pq}\right)^{-1}-\frac{\delta_{pq}}{2}\right]\left(D_q g_q^{eq}-G_q\right)\\
		=&\left[({\boldsymbol{S}^{g}_{2}})^{-1}-\frac{\mathbf{I}}{2}\right]\sum_q \mathbf{c}_{q\alpha} \mathbf{c}_{q\beta}(D_q g_q^{eq}-G_q)-({\boldsymbol{S}^{g}_{2}})^{-1}\delta_{\alpha\beta}\frac{\sum_q \mathbf{c}_q \cdot\mathbf{c}_q(D_q g_q^{eq}-G_q)}{d}\\
		&+({s^{g}_{0}})^{-1}c_s^2\delta_{\alpha\beta}\sum_q(D_q g_q^{eq}-G_q),
	\end{aligned}
\end{equation}
then we mark $\Pi_{\alpha\beta}^1=\sum_p\textbf{c}_{p\alpha}\textbf{c}_{p\beta}(D_p g_p^{eq}-G_p)$, $\Pi^2=\sum_p\textbf{c}_p\cdot\textbf{c}_p(D_p g_p^{eq}-G_p)$, 
and $\Pi^3=\sum_p(D_p g_p^{eq}-G_p)$, and they can be calculated by 
\begin{equation}
	\begin{aligned}
		\Pi_{\alpha\beta}^1= &\partial_t\left(\frac{\phi{m}_\alpha^{ C} u_{\beta}+\phi u_{\alpha} m_{\beta}^{C}}{2}\right)+c_s^2\left[\nabla_\alpha (\phi\rho u_{\beta})+\nabla_{\beta}(\phi \rho u_\alpha)\right]+\nabla_{\gamma}(\phi\rho\bar{\delta}_{\alpha\beta\gamma\eta}^{(4)}u_{\eta})\\
		&+c_s^2\phi u_{\gamma} \nabla_{\gamma} \rho \delta_{\alpha\beta}-c_s^2\phi u_{\gamma} \nabla_{\gamma} \rho \delta_{\alpha\beta}-{M}^{2G}_{\alpha\beta}, \\
		\Pi^2=&\partial_t\left(\phi{m}_\alpha^{ C} u_{\alpha}\right)+2c_s^2\nabla_\alpha (\phi\rho {u}_\alpha)+\nabla_{\gamma}(\phi\rho\bar{\delta}_{\alpha\alpha\gamma\eta}^{(4)}u_{\eta})+dc_s^2\phi {u}_\alpha\nabla_\alpha \rho\\
		&-dc_s^2\phi {u}_\alpha\nabla_\alpha \rho  -{M}^{2G}_{\alpha\alpha},\\
		\Pi^3=& \partial_{t}\phi\rho_0+	\nabla\cdot(\phi\rho\textbf{u})-\phi\textbf{u}\cdot\nabla\rho.\\
	\end{aligned}
	\label{eq:B10}
\end{equation}
By taking the following expression of $\mathbf{M}^{2G}$,
\begin{equation}
	\mathbf{M}^{2G}=\partial_t\left(\frac{\phi\mathbf{m}^{ C} \mathbf{u}+\phi\mathbf{u m}^{ C}}{2}\right)+c_s^2\left[\mathbf{u} \nabla(\phi \rho)+\nabla( \phi\rho) \mathbf{u}\right]+(c^2-3c_s^2)\mathbf{u} \cdot\nabla(\phi\rho) \mathbf{I},
\end{equation}
and substituting Eq. (\ref{eq:B10}) into Eq. (\ref{eq:B11}), and this equation can be rewritten as
\begin{equation}
	\partial_{t}(\phi\rho\textbf{u})+\nabla\cdot\left(\phi\rho\textbf{u}\textbf{u}+\frac{\phi\textbf{m}^{C}\textbf{u}+\phi\textbf{u}\textbf{m}^{C}}{2}+\phi p\textbf{I}\right)=
	\nabla\cdot \tau+\textbf{F}+O(\delta t^2),
\end{equation}
where 
\begin{equation}
	\tau=\delta_t\phi\rho\left[({\boldsymbol{S}^{g}_{2}})^{-1}-\frac{\textbf{I}}{2}\right]\left[c_s^2(\nabla\textbf{u}+(\nabla\textbf{u})^{T})+\nabla\cdot(\bar{\delta}^{(4)}\cdot \textbf{u})\right].
\end{equation}
Thus the macroscopic incompressible DD-NS equations can be recovered at the order of $O(\delta t^2)$ with 
\begin{equation}
	\nu  =  \left(\frac{1}{s^g_{21}}-\frac{1}{2}\right)\frac{c^2-c_s^2}{2}\delta t,\quad \nu = \left(\frac{1}{s^g_{22}}-\frac{1}{2}\right)c_s^2\delta t.
\end{equation}

\section{The computation of Pressure}\label{sec:appendixC}
\setcounter{equation}{0}
\renewcommand\theequation{C.\arabic{equation}}
Now let us focus on the computation of pressure $P$. From Eq. (\ref{eq:B4}), one can obtain the zeroth direction of $g_p^{ne}$ ,
\begin{equation}
	\begin{aligned}
		g_0^{ne}  =&-\delta t \partial_t\left[(\Lambda^{g}_{0 q})^{-1} g_q^{eq}\right]+\delta t\left(\Lambda_{0q}^{-1} G_q-\frac{G_0}{2}\right)+O\left(\delta t^2\right) \\
		=&\delta t \frac{K}{c^2 s^g_{21}} \partial_t(\phi \mathbf{m} \cdot \mathbf{u})+\delta t\left[\frac{K\left(s^g_{21}-2\right)}{2 c^2 s^g_{21}} \partial_t(\phi \mathbf{m} \cdot \mathbf{u})+\frac{H}{s^g_{21}} \mathbf{u} \cdot \nabla( \phi \rho)\right.\\
		&\left.+\frac{K(1-d_0)(2-s^g_0)}{2 s^{g}_0} \phi \mathbf{u} \cdot \nabla \rho\right] .\\
	\end{aligned}
\end{equation}
According to above equation and $g_0=g_0^{\mathrm{eq}}+g_0^{n e}$, we have
\begin{equation}
	\begin{aligned}
		\frac{1-\omega_0}{c_s^2} \phi P  =&\phi\rho_0-\left(g_0-g_0^{n e}\right)+s_0 \\
		=&\sum_{p \neq 0} g_p+\left[\frac{1}{2}+\frac{K(1-d_0)(2-s^{g}_0)}{2s^{g}_0}\right] \delta t \phi\mathbf{u} \cdot \nabla \rho+ \delta t\frac{K}{2c^2} \partial_t\left(\phi\mathbf{m}^{\phi C} \cdot \mathbf{u}\right)\\
		&+\delta t\frac{H}{s^g_{21}} \mathbf{u} \cdot \nabla( \phi\rho)+s_0+O\left(\delta t^2\right) .\\
	\end{aligned}
\end{equation}
Neglecting the truncation error term $O\left(\delta t^2\right)$, one can obtain the computational scheme (\ref{eq:19}) for pressure $P$.

\section{Analytical solution of the equilibrium shape of a quaternary-phase compound drop}\label{sec:appendixD}
\setcounter{equation}{0}
\renewcommand\theequation{D.\arabic{equation}}
\begin{figure}[H]{
		\centering
		\includegraphics[width=2in]{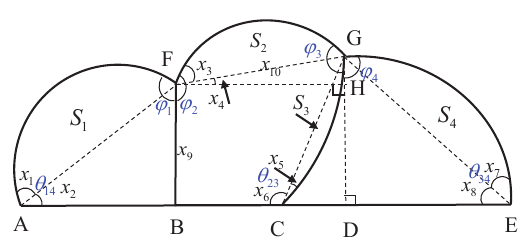}
		\caption{The configuration for the equilibrium shape of a quaternary-phase compound drop.}}
	\label{fig:10}
\end{figure}
The analytical solution for the shape of a compound drop on the idea wall can be derived based on the mass conservation and force balance. Now we give some details on how to derive the analytical solution. Under the assumption of $\theta_{12}=90^{\circ}$, the equilibrium shape of the quaternary-phase compound drop is shown in Fig. \ref{fig:10}, where some variables about the angles are denoted by $x_{1-8}$, and the lengths of $\overline{FB}$ and  $\overline{FG}$ are set as $x_{9,10}$. According to the definition of the contact angle, we can easily obtain the following relations,
\begin{equation}
	\begin{aligned}
		x_1+x_2=\theta_{14}, \quad x_5+x_6=\theta_{23}, \quad x_7+x_8=\theta_{34}, \\		
		x_1 -x_2+\frac{\pi}{2}= \varphi_1, \quad x_3+x_4+\frac{\pi}{2}=\varphi_2, \\
		x_3+x_5+\pi-x_4-x_6=\varphi_3, \quad x_6+x_7-x_5-x_8=\varphi_4.
	\end{aligned}
	\label{D.1}
\end{equation}
When the initial radius $R$ of a compound droplet is given, the initial areas of three droplets can be obtained by $S_{f{1,2,3}}$, then we can derive the expression of the areas of the circular segments:
\begin{equation}
	\begin{aligned}
		&S_1=\frac{x_1-\sin x_1 \cos x_1}{4 \sin ^2 x_1 \cos ^2 x_1} x_9^2,\\
		&S_2=\frac{x_3-\sin x_3 \cos x_3}{4 \sin ^2 x_3} x_{10}^2,\\
		& S_3=\frac{x_5-\sin x_5 \cos x_5}{4 \sin ^2 x_5 \sin ^2 \left(\pi-x_6\right)}\left(x_9+x_{10} \sin x_4\right)^2,\\ 
		&S_4=\frac{x_7-\sin x_7 \cos x_7}{4 \sin ^2 x_7 \sin ^2 x_8}\left(x_9+x_{10} \sin x_4\right)^2.\\
	\end{aligned}
\end{equation}
In addition, the areas of the triangles $\triangle AFB$, $\triangle CGE$, $\triangle CGE$ and the trapezoid $\square BDGF$ can be determined by
\begin{equation}
	\begin{aligned}
		&S_{\triangle AFB}=\frac{x_9^2 }{2\tan x_2}, \\
		&S_{\triangle DGE}=\frac{\left(x_9+x_{10} \sin x_4\right)^2}{2\tan x_8},\\
		&S_{\triangle CGD}=\frac{\left(x_9+x_{10} \sin x_4\right)^2}{2\tan \left(\pi-x_6\right)},\\
		&S_{\square BDGF}=\frac{2x_9+x_{10} \sin x_4 }{2}x_{10} \cos x_4.\\
	\end{aligned}
\end{equation}
Therefore, the following equivalence relations of the areas of the compound droplet hold
\begin{equation}
	\begin{aligned}
		& S_{f_1}=	S_{1}+ 	S_{\triangle  AFB},\\
		& S_{f_2}= S_{2}+ S_{3}+ S_{\square BDGF}-	S_{\triangle CGD},\\
		& S_{f_3}=S_{4}- S_{3}+S_{\triangle CGD}+	S_{\triangle EGD}.\\
	\end{aligned}
	\label{D.4}
\end{equation}
If the physical quantities $\theta_{14}$, $\theta_{23}$, $\theta_{34}$, $\varphi_1$, $\varphi_2$, $\varphi_3$ and $\varphi_4$ are fixed, Eqs. (\ref{D.1}) and (\ref{D.4}) constitute a closed system for the variables $x_{1-10}$, and their solutions can thus be uniquely determined. When these variables are given, the analytical shape of the compound drop at equilibrium can also be uniquely determined. Then the spreading lengths of phases 1, 2, 3 denoted by $L_1$, $L_2$, and $L_3$ can be expressed as
\begin{equation}
	\begin{aligned}
		&L_1=\overline{A B}= \frac{x_9}{\tan x_2},\\
		&L_2=\overline{BC}={x_{10}}\cos x_4- \frac{x_9+x_{10}\sin x_4}{\tan(\pi-x_6)},\\
		&L_3=\overline{CE}={x_{10}}\cos x_4- \frac{x_9+x_{10}\sin x_4}{\tan x_8 +\tan(\pi-x_6)}.\\
	\end{aligned}
\end{equation}
The solutions of these equations cannot be given explicitly, and we choose to numerically solve them by using the Newton iteration method and show some results in Table \ref{table3}.	
%\bibliographystyle{elsarticle-num}
%\bibliography{rlbm}


\bibliographystyle{siamplain}
\bibliography{references}
\end{document}